\documentclass[aps,prl,reprint,superscriptaddress,longbibliography]{revtex4-2}

\usepackage{microtype}
\usepackage{amsmath,amssymb}
\usepackage{graphicx}
\usepackage{subcaption} 

\usepackage{hyperref}
\usepackage{xcolor}
\usepackage{quantikz}
\hypersetup{hidelinks}

\usepackage{url}

\makeatletter
\def\prl@sectionstyle{%
  \@startsection{section}{1}{\parindent}%
  {1.2ex}
  {-1sp}%
  {\normalfont\itshape}%
}

\let\prl@origsection\section
\def\section{\@ifstar\prl@sectionstar\prl@sectionnostar}
\def\prl@sectionnostar#1{\prl@sectionstyle{#1\textemdash}}
\def\prl@sectionstar#1{\prl@sectionstyle*{#1\textemdash}}
\makeatother

\begin{document}

\title{Hardware-Aware Realization of Deterministic Quantum-Walk Search on Complete Bipartite Graphs}

\newcommand{\affA}{Hefei National Research Center for Physical Sciences at the Microscale and School of Physical Sciences, University of Science and Technology of China, Hefei 230026, China}

\newcommand{\affB}{Shanghai Research Center for Quantum Science and CAS Center for Excellence in Quantum Information and Quantum Physics, University of Science and Technology of China, Shanghai 201315, China}

\newcommand{\affC}{Hefei National Laboratory, University of Science and Technology of China, Hefei 230088, China}

\newcommand{\affD}{State Key Lab of Processors, Institute of Computing Technology, Chinese Academy of Sciences, Beijing 100190, China}

\newcommand{\affE}{School of Computer Science and Technology, University of Chinese Academy of Sciences, Beijing 100049, China}


\author{Tao Rong}
\thanks{These authors contributed equally to this work.}
\affiliation{\affA}
\affiliation{\affB}
\affiliation{\affC}

\author{Meng Li}
\thanks{These authors contributed equally to this work.}
\affiliation{\affD}
\affiliation{\affE}

\author{Xin Tong}
\thanks{These authors contributed equally to this work.}
\affiliation{\affD}

\author{Jianbin Cai}
\affiliation{\affA}
\affiliation{\affB}
\affiliation{\affC}

\author{Yu Li}
\affiliation{\affD}
\affiliation{\affE}

\author{Naibin Zhou}
\affiliation{\affA}
\affiliation{\affB}
\affiliation{\affC}

\author{Shaowei Li}
\affiliation{\affA}
\affiliation{\affB}
\affiliation{\affC}

\author{Sirui Cao}
\affiliation{\affA}
\affiliation{\affB}
\affiliation{\affC}

\author{Yuan Li}
\affiliation{\affA}
\affiliation{\affB}
\affiliation{\affC}

\author{Yu Xu}
\affiliation{\affA}
\affiliation{\affB}
\affiliation{\affC}

\author{Fusheng Chen}
\affiliation{\affA}
\affiliation{\affB}
\affiliation{\affC}

\author{Ming Gong}
\affiliation{\affA}
\affiliation{\affB}
\affiliation{\affC}

\author{Xiaoming Sun}
\email{sunxiaoming@ict.ac.cn}
\affiliation{\affD}
\affiliation{\affE}

\author{Chen Zha}
\email{chenzha@hfnl.cn}
\affiliation{\affA}
\affiliation{\affB}
\affiliation{\affC}


\begin{abstract}

Complete bipartite graphs offer a canonical platform for deterministic quantum search with quadratic speedup. We report the first experimental realization of deterministic discrete-time quantum-walk search over complete bipartite graphs, enabled by high-fidelity operations and a dynamical-phase compensation scheme, using a digital gate-based processor rather than an evolution-based platform. 
By compiling the search circuits to the native gate set and square-lattice topology, we implement full-space encoding to realize instances of graph sizes $N=4$ and $8$, with success probabilities of 97.6\% and 70.8\% for finding the marked vertex, respectively. We also realize resource-efficient exact instances up to size $N=16$, the largest achieving 66\% success probability. Separately, a reduced invariant-subspace encoding preserves the dynamics of the deterministic search algorithm on two qubits for nominal problem sizes up to $N=256$, with the success probability approaching 50\%.  
Resource analyses of logical, hardware-compiled, and fault-tolerant-gate-set-synthesized circuits show that circuit depth and size scale as $\sqrt{N}$, with hardware overheads entering only as constant prefactors.
For application-scale graphs ($N \sim 10^{5}$), an effective gate-fidelity model requires per-gate infidelity near $10^{-7}$. 
These results establish a concrete algorithm-to-hardware pathway for deterministic structured quantum search on gate-based quantum processors.
\end{abstract}

\maketitle
\raggedbottom

\section{Introduction}
Complete bipartite graphs model two-mode interaction systems across recommendation, biological, and quantum spin networks, where the absence of intra-class edges imposes a spectral symmetry shaping all dynamical processes, while bipartition and marked-set asymmetry yield richer interference dynamics that better reflect structured real-world systems~\cite{cambie2025packing, bara2025independent}. 
Although complete bipartite graphs are highly symmetric, bipartite quadratic assignment problem-type problems remain NP-hard on them~\cite{punnen2016bipartite}. This motivates our study of search algorithms for complete bipartite graphs.

Quantum walks, realized in continuous and discrete forms across platforms~\cite{ryan2005experimental, lu2010experimental, grafe2016integrated, neves2018photonic, yan2019strongly, gong2021quantum, zhou2025demonstration, dubovitskii2026experimental, razzoli2024efficient, shakeel2020efficient, georgopoulos2021comparison, wing2023circuit, wadhia2024cycle}, achieve quadratic and even exponential speedups for certain structured search problems~\cite{Aharonov2001QuantumWalks, childs2003exponential, Childs2004SpatialSearch, Chakraborty2016OptimalSpatialSearch, Ambainis2020MarkedVertices, apers2022quadratic, peng2024search, peng2024lackadaisical}. Yet their success probabilities oscillate, preventing deterministic retrieval, which motivates schemes concentrating all probability at a prescribed time. 
Such ideas are well developed for Grover-type search~\cite{Hoyer2000ArbitraryPhases, Long2001ZeroFailureGrover} and extended to several continuous-time quantum-walk (CTQW) models~\cite{Marsh2021DeterministicSpatialSearch, Qu2022DeterministicStarGraphs, wang2025unifying}. However, CTQW's Hamiltonian formulation aligns less naturally with gate-based, fault-tolerant computing, whereas discrete-time quantum-walk (DTQW) maps directly to circuits.  
DTQW achieves deterministic search on complete graphs~\cite{Hoyer2000ArbitraryPhases, Long2001ZeroFailureGrover}, but this does not transfer to complete bipartite graphs, where bipartition and marked-set asymmetry break the simple two-state dynamics.
Building on this, a recent deterministic DTQW search algorithm~\cite{Peng2024DeterministicDTQW} overcomes the obstacle using a generalized Grover coin and a tunable phase oracle, mapping the dynamics onto an effective four-state (or eight-state) model and concentrating all probability on the marked vertex at a prescribed time, thereby achieving unit success probability with quadratic speedup.

Nevertheless, translating this ideal algorithm into executable circuits on gate-based hardware is nontrivial, requiring edge-space encoding, oracle and coin synthesis, conditional-shift implementation, hardware compilation and fidelity analysis.

We validate this full pipeline on the gate-based hardware~\cite{jiang2025generation95qubitgenuineentanglement}, compiling the resulting circuits to the native gate set and square-lattice topology to analyze how circuit depth, gate count, and required gate fidelity scale with graph size.

We then report the first experimental realization of quantum-walk search on a complete bipartite graph, with a success probability exceeding the same-query classical baseline~\cite{pokharel2024better} at graph size $N = 8$, and introduce a reduced invariant-subspace encoding that extends the accessible range on two qubits for the scalability benchmark. Together, these results identify circuit depth and accumulated gate errors—not Hilbert-space dimension alone—as the dominant scaling bottleneck. This work establishes an explicit algorithm-to-hardware pathway for structured quantum search on future processors.

\section{Algorithm Setting and Circuit Construction}

The deterministic search protocol is naturally formulated as a discrete-time quantum walk on the edge space of a complete bipartite graph. Following Ref.~\cite{Peng2024DeterministicDTQW}, we consider $G(V,E)$ as a complete bipartite graph with vertex partition $V=V_1\cup V_2$, where $|V_1|=N_1$ and $|V_2|=N_2$. The walk acts on the edge Hilbert space spanned by $\{\ket{v,u}:(v,u)\in E\}$.

\begin{figure*}[t]
    \centering
    \includegraphics{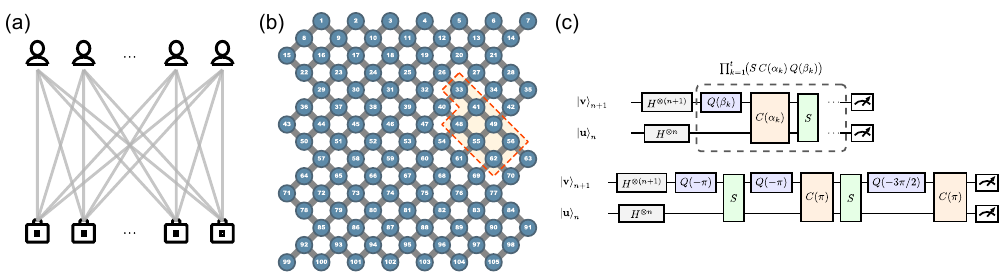}
    \caption{Overview of the experimental platform and algorithm implementation. (a) Schematic representation of embedding an NRMP problem into a complete bipartite graph, where the top nodes represent all active applicants and the bottom nodes represent all available positions. (b) Square-lattice topology and physical qubit connectivity of the quantum processor used in this work. (c) Full algorithmic quantum circuits for the general and special cases. }
    \label{fig:overall_setup}
\end{figure*}

The one-step walk search operator is
$U(\alpha,\beta)=SC(\alpha)Q(\beta)$, with the coin operator $C(\alpha)\ket{v,u}=\ket{v}\otimes G_v(\alpha)\ket{u}$ and $G_v(\alpha)=(1-\mathrm{e}^{-\mathrm{i}\alpha})\ket{s_v}\bra{s_v}-I$, where $\ket{s_v}=\frac{1}{\sqrt{d_v}}\sum_{(v,u)\in E}\ket{u}$ and $d_v$ represents the degree of vertex $v$.
The conditional shift satisfies $S\ket{v,u}=\ket{u,v}$, and the oracle acts as
{
\setlength{\abovedisplayskip}{6pt}
\setlength{\belowdisplayskip}{6pt}
\begin{align*}
Q(\beta)\ket{v,u}=
\begin{cases}
    \mathrm{e}^{\mathrm{i}\beta}\ket{v,u}, & v\in M,\\
    \ket{v,u}, & v\notin M,
\end{cases}
\end{align*}
}
for marked set $M\subset V$. 

With $M\subset V_1$ and $|M|=n_1>0$ known, the following four states span a $U(\alpha,\beta)$-invariant subspace:
{
\setlength{\abovedisplayskip}{6pt}
\setlength{\belowdisplayskip}{6pt}
\allowdisplaybreaks[1]

\begin{align*}
\ket{\phi_1}&=\frac{1}{\sqrt{n_1N_2}}\sum_{v\in M,\,u\in V_2}\ket{v,u},\\[-6pt]
\ket{\phi_2}&=\frac{1}{\sqrt{(N_1-n_1)N_2}}\sum_{v\in V_1\backslash M,\,u\in V_2}\ket{v,u},\\[-6pt]
\ket{\phi_3}&=\frac{1}{\sqrt{n_1N_2}}\sum_{v\in V_2,\,u\in M}\ket{v,u},\\[-6pt]
\ket{\phi_4}&=\frac{1}{\sqrt{(N_1-n_1)N_2}}\sum_{v\in V_2,\,u\in V_1\backslash M}\ket{v,u}.
\end{align*}
}
With appropriate parameters $\{\alpha_j,\beta_j\}$, the initial state $\ket{\psi(0)}=\frac{1}{\sqrt{2N_1N_2}}\sum_{v\in V,\,(v,u)\in E}\ket{v,u}$ evolves to unit success probability after $O(\sqrt{N_1/n_1})$ walk steps~\cite{Peng2024DeterministicDTQW}.

As a practical motivation, the National Resident Matching Program (NRMP)~\cite{roth1999redesign,nrmp2026report} maps naturally onto a complete bipartite graph, with applicants and residency positions as the two vertex sets, as shown in Fig.~\ref{fig:overall_setup}(a). Locating a marked applicant or position is therefore a structured-search task on this graph. We assess the viability of the deterministic DTQW protocol by estimating the resources required after hardware compilation and running the hardware-compiled circuits of Fig.~\ref{fig:overall_setup}(c) on the quantum experiment platform depicted in Fig.~\ref{fig:overall_setup}(b), including both the general full-space encoding and the exact $n_1/N_1=1/4$ resource-efficient instance.
\begin{figure*}[t]
    \centering
    \includegraphics[width=\textwidth]{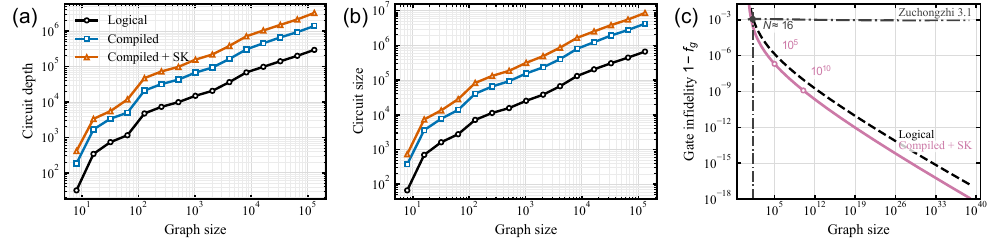}
    \caption{Resource scaling and gate-fidelity requirements for the general complete-bipartite graph search circuit. (a) Circuit depth versus graph size for the logical circuit, the hardware-compiled circuit mapped to the square-lattice topology and native gate set, and the SK-synthesized circuit obtained by subsequent Solovay--Kitaev (SK) synthesis. (b) Corresponding circuit size versus graph size. (c) Required per-gate infidelity, $1-f_g$, versus graph size under $F_{\mathrm{tot}}\ge 0.5$, estimated from fitted gate-count models for the logical and SK-synthesized circuits; the vertical dashed line indicates the current \textit{Zuchongzhi 3.1} graph-size limit estimated from calibrated one- and two-qubit gate fidelities using the effective gate-fidelity model.}
    \label{fig:sim-summary}
\end{figure*}

\section{Resource scaling and gate-fidelity requirements}

To assess the practical feasibility of the deterministic DTQW search algorithm, we analyze how its resource cost grows with graph size after realistic hardware compilation. We focus on the general full-space construction, since it directly realizes the original search problem and therefore governs the asymptotic cost. The special case, by contrast, has a fixed circuit structure and does not probe the same bottleneck. For this analysis we adopt the balanced encoding $N_1=N_2=2^{n}$, giving a total size $N=2^{n+1}$, where $n$ is the bit length of one partition and is distinct from the marked-set size $n_1=|M|$. The vertex/edge-register encoding and the oracle, coin, and shift definitions are detailed in the Supplemental Material~\cite{SM}.

To distinguish algorithmic scaling from hardware overhead, we compare three circuit representations up to $N\simeq1.3\times10^{5}$ ($n=16$) in Fig.~\ref{fig:sim-summary}(a,b). The logical circuit has no connectivity constraint; the hardware-compiled circuit uses square-lattice connectivity and the native gate set $\{H,R_z,\mathrm{CZ}\}$; and the SK-synthesized circuit further applies Solovay--Kitaev (SK) synthesis~\cite{Dawson2006SolovayKitaev} over $\{H,T,T^{\dagger},Z\}$, retaining CZ gates. The last stage projects resources for fault-tolerant implementations, excluding error-correction overhead; experiments use hardware-compiled circuits without SK synthesis. Multi-controlled phase rotations dominate each walk step, requiring $O(\log n)$ depth and $O(n)$ gates~\cite{Nie2024QuantumCircuitToffoli}. With $O(\sqrt{2^{n}})$ search iterations, both total depth and size scale as $\sqrt{N}$ up to logarithmic factors. Over the investigated range, hardware compilation increases depth by a factor of $4$--$5$ and size by $5$--$6$, while SK synthesis adds a further factor of about $2$ to both, preserving the leading $\sqrt{N}$ scaling.

Because a direct success-rate estimation becomes intractable at large $N$, we use the total circuit fidelity $F_{\mathrm{tot}}=f_g^{N_g}$ as a scalable proxy for the hardware requirement, where $f_g$ is the effective elementary-gate fidelity and $N_g$ is a fitted gate count obeying $\log N_g = an + b\log n + c$. Taking $F_{\mathrm{tot}}\ge0.5$ as an order-unity criterion, the comparison in Fig.~\ref{fig:sim-summary}(c) shows that hardware compilation and SK synthesis tighten the required per-gate infidelity by roughly an order of magnitude relative to the logical estimate. Applying this model to the calibrated \textit{Zuchongzhi 3.1} fidelities places the current device at only a small graph size, as marked by the vertical dashed line in Fig.~\ref{fig:sim-summary}(c), confirming that present hardware only supports proof-of-principle full-space demonstrations. Reaching application scale is far more demanding. A graph size $N\sim10^{5}$, matching the latest NRMP instance~\cite{roth1999redesign, nrmp2026report}, already requires a per-gate infidelity near $10^{-7}$, and $N\sim10^{10}$ requires the $10^{-9}$ level, setting concrete logical-gate performance targets for fault-tolerant implementations.

\begin{figure*}
    \centering
    \includegraphics{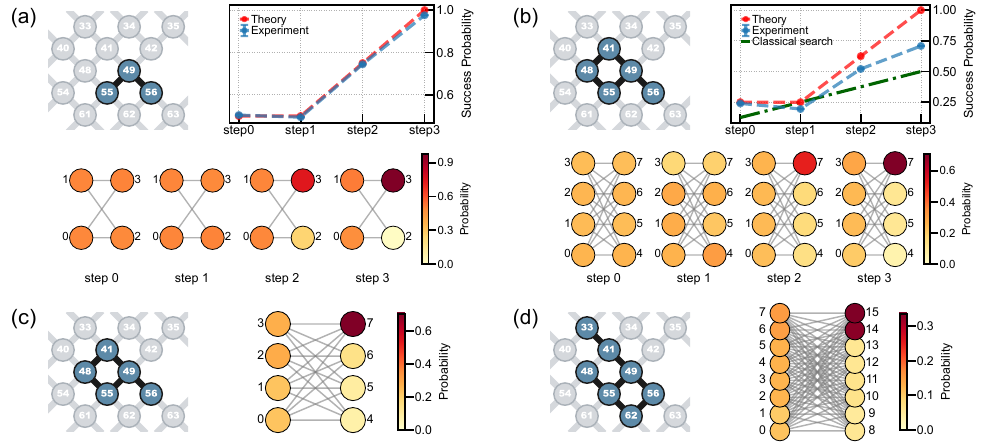}
    \caption{\label{fig:prob-plot-gen}
     Experimental realizations of deterministic DTQW search on complete bipartite graphs. Panels (a) and (b) correspond to the general case with graph sizes $N=4$ and $N=8$, respectively, and panels (c) and (d) correspond to the special case with $N=8$ and $N=16$, respectively. The processor layouts indicate the qubits and couplers of the \textit{Zuchongzhi 3.1} processor used in each experiment. The probability plots show the target-vertex success probability versus walk step; dashed lines denote theoretical predictions, symbols denote experimental data, and error bars represent one standard deviation estimated from $10^5$ measurement shots. The dash-dotted line in (b) denotes the classical success probability $P_{\mathrm{cl}}$. The colored complete-bipartite graphs show the vertex probability distributions at successive steps for the general case and at the final step for the special case. }
\end{figure*}

\section{Experimental Results}
Following the circuit-level resource analysis, we implement both the general full-space circuit and the resource-efficient special-case circuit on the \textit{Zuchongzhi 3.1} superconducting quantum processor. The logical walk operator $U(\alpha,\beta)$ is compiled into the native gate set consisting of single-qubit $\pi/2$ rotations and CZ gates, with circuit depth increasing with graph size. We experimentally realize circuits involving up to seven qubits, corresponding to a graph size of $N = 16$. To mitigate dephasing during long idle periods, dynamical-decoupling pulses are inserted throughout circuit execution. We further calibrate and compensate the dynamical phases induced by transient effects of the CZ control pulses using virtual-Z corrections, as detailed in the Supplemental Material \cite{SM}.

In the general case, we performed search experiments for graph sizes $N=4$ and $N=8$, with the results shown in Fig.~\ref{fig:prob-plot-gen}(a) and (b), respectively. In both instances, the success probability increases with the number of walk steps, reaching final values of 97.6\% and 70.8\%, respectively. For the $N=8$ case, the measured success probability exhibits a sharp increase after $t=1$ and surpasses  the classical success probability $P_{\mathrm{cl}}=\frac{t+1}{N}$~\cite{pokharel2024better} in the final two steps, as shown in Fig.~\ref{fig:prob-plot-gen}(b), where $t$ denotes the number of oracle queries and $N$ is the graph size. These results demonstrate the effective accumulation of probability amplitude on the target vertex and provide a proof-of-principle experimental validation of deterministic full-space search on complete bipartite graphs, whose ideal success probability reaches unity. Scaling the general construction to $N=16$ already requires 743 CZ layers, exceeding the reliable depth of the present control hardware. This implementation limit is consistent with the resource analysis of Fig.~\ref{fig:sim-summary}, where the hardware-compiled circuit depth is identified as the dominant scaling overhead. The corresponding experiment was therefore not performed. This limitation identifies circuit depth and accumulated gate errors as the main bottlenecks for extending the direct full-space implementation to larger graphs.

For the special case $n_1/N_1=1/4$, the logical circuit consists of an initial layer of Hadamard gates on the vertex and edge registers, followed by the fixed sequence
$Q(-\pi), S, Q(-\pi), C(\pi), S, Q(-3\pi/2), C(\pi)$,
as shown in Fig.~\ref{fig:overall_setup}(c).
We demonstrated search experiments for graph sizes $N=8$ and $16$, yielding final success probabilities of 70.3\% and 66.0\%, respectively, as shown in Fig.~\ref{fig:prob-plot-gen}(c) and (d). Notably, the success probability in this case does not decrease significantly with increasing graph size. Together with the resource analysis of the general case, these results indicate that the principal limitation to scaling arises not merely from the growth of Hilbert-space dimension, but more critically from the accompanying increase in circuit depth. By substantially mitigating this depth overhead, the special-case construction provides a promising route toward scalable implementations of deterministic DTQW search on larger gate-based quantum processors.

\begin{figure*}
    \centering
    \includegraphics{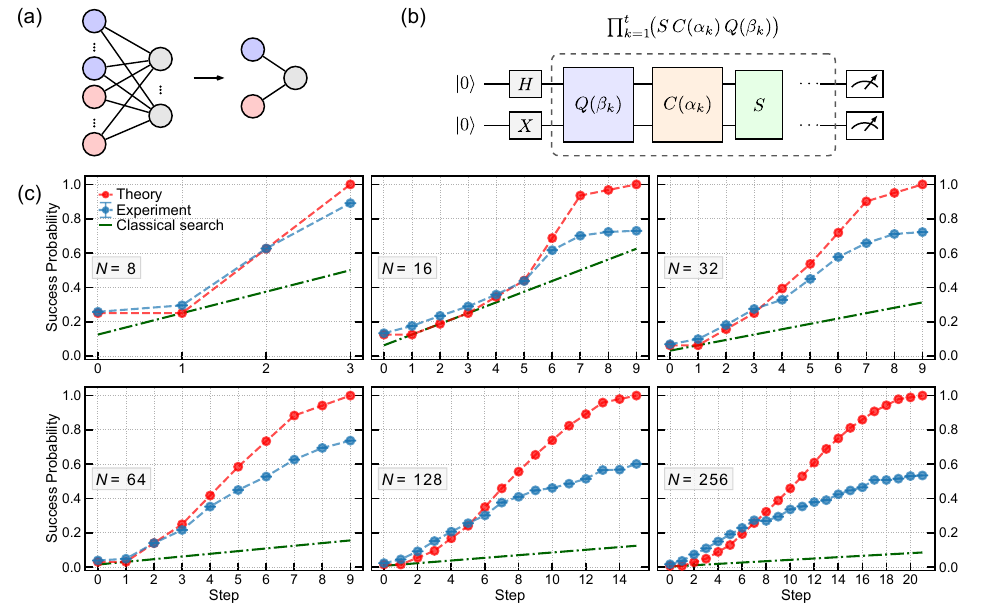}
    \caption{Reduced-subspace demonstration of scalable complete-bipartite DTQW search dynamics. 
    (a) Subspace reduction via class aggregation from the original complete bipartite graph (left) to an effective three-vertex representation (right). 
    (b) Quantum circuit for the reduced-subspace implementation. 
    (c) Target-vertex success probability versus walk step for graph sizes $N=2^k$ ($k=3,\dots,8$), as indicated by the gray labels. Red dashed lines denote theoretical predictions, blue symbols denote experimental data, and green dash-dotted lines denote classical success probability $P_{\mathrm{cl}}$. Error bars represent one standard deviation estimated from $10^5$ measurement shots.}
    \label{fig:prob-plot-sub}
\end{figure*}

\section{Reduced-subspace scalability benchmark}
To further benchmark experimental scalability beyond the graph sizes accessible to the direct full-space encoding, we introduce the reduced-subspace construction illustrated in Fig.~\ref{fig:prob-plot-sub}(a) as an effective-size probe of the target walk dynamics. Although this construction does not encode all graph vertices explicitly, it faithfully reproduces the effective search dynamics of the original
 walk within its invariant subspace. It therefore provides a controlled way to test how the deterministic interference mechanism behaves as the nominal graph size and the
 corresponding walk schedule increase.

In this setting, the walk is restricted to the four-dimensional invariant subspace spanned by $\{\ket{\phi_1},\ket{\phi_2},\ket{\phi_3},\ket{\phi_4}\}$ introduced above, so the evolution can be represented with two qubits, independent of the graph size. This reduced construction provides a compact setting for benchmarking the target walk dynamics and diagnosing hardware errors. Following the parameter schedule of Ref.~\cite{Peng2024DeterministicDTQW}, we construct a state-preparation block followed by a sequence of walk steps implementing $U(\alpha_j,\beta_j)=SC(\alpha_j)Q(\beta_j)$ on the reduced space. The corresponding logical circuit is shown in Fig.~\ref{fig:prob-plot-sub}(b).

In the subspace case, we clearly observe the search success probability rising as the number of walk steps increases for nominal graph sizes from $N=2^3$ to $N=2^8$, as shown in Fig.~\ref{fig:prob-plot-sub}(c). Specifically, the success probability for $N = 8$ reaches 90\%, and for $N=16$, $32$, and $64$, it remains around 80\%, showing that the deterministic-search population transfer is preserved over a range of effective graph sizes. At larger nominal sizes, the deeper walk schedules lead to a gradual reduction of the peak probability, reaching about 60\% for $N=128$ and approaching 50\% for $N=256$, where decoherence drives the output close to a mixed state. These results indicate that the protocol is experimentally viable at the circuit level and that its effective search dynamics can be preserved over larger nominal graph sizes when the full encoding overhead is removed.

\section{Conclusion}
We have demonstrated deterministic discrete-time quantum-walk search on complete bipartite graphs using a gate-based digital processor. By combining full-space and invariant-subspace encodings, we preserve exact search dynamics across a wide range of effective graph sizes, while our resource analysis identifies accumulated gate errors—rather than problem size alone—as the central scaling bottleneck. This establishes a concrete algorithm-to-hardware pathway for structured quantum search and motivates low-error gate synthesis and error mitigation for application-scale implementations.

\section{Acknowledgments}
This research was supported by the Quantum Science and Technology–National Science and Technology Major Project (Grant No. 2024ZD0300500); the Quantum Science and Technology-National Science and Technology Major Project (Grant No. 2021ZD0300200); the National Natural Science Foundation of China (Grant Nos. 92476203, 92476001, 62325210, and 62301531); the New Cornerstone Science Foundation through the XPLORER PRIZE; the Jinan Science and Technology Bureau and the Jinan High-Tech Zone Management Committee; the Cultivation Project of the Shanghai Research Center for Quantum Sciences (Grant No. LZPY2024); the Beijing Natural Science Foundation (Grant No. 4252012); and the Taishan Scholars Program.


%


\begin{thebibliography}{40}%
\makeatletter
\providecommand \@ifxundefined [1]{%
 \@ifx{#1\undefined}
}%
\providecommand \@ifnum [1]{%
 \ifnum #1\expandafter \@firstoftwo
 \else \expandafter \@secondoftwo
 \fi
}%
\providecommand \@ifx [1]{%
 \ifx #1\expandafter \@firstoftwo
 \else \expandafter \@secondoftwo
 \fi
}%
\providecommand \natexlab [1]{#1}%
\providecommand \enquote  [1]{``#1''}%
\providecommand \bibnamefont  [1]{#1}%
\providecommand \bibfnamefont [1]{#1}%
\providecommand \citenamefont [1]{#1}%
\providecommand \href@noop [0]{\@secondoftwo}%
\providecommand \href [0]{\begingroup \@sanitize@url \@href}%
\providecommand \@href[1]{\@@startlink{#1}\@@href}%
\providecommand \@@href[1]{\endgroup#1\@@endlink}%
\providecommand \@sanitize@url [0]{\catcode `\\12\catcode `\$12\catcode
  `\&12\catcode `\#12\catcode `\^12\catcode `\_12\catcode `\%12\relax}%
\providecommand \@@startlink[1]{}%
\providecommand \@@endlink[0]{}%
\providecommand \url  [0]{\begingroup\@sanitize@url \@url }%
\providecommand \@url [1]{\endgroup\@href {#1}{\urlprefix }}%
\providecommand \urlprefix  [0]{URL }%
\providecommand \Eprint [0]{\href }%
\providecommand \doibase [0]{https://doi.org/}%
\providecommand \selectlanguage [0]{\@gobble}%
\providecommand \bibinfo  [0]{\@secondoftwo}%
\providecommand \bibfield  [0]{\@secondoftwo}%
\providecommand \translation [1]{[#1]}%
\providecommand \BibitemOpen [0]{}%
\providecommand \bibitemStop [0]{}%
\providecommand \bibitemNoStop [0]{.\EOS\space}%
\providecommand \EOS [0]{\spacefactor3000\relax}%
\providecommand \BibitemShut  [1]{\csname bibitem#1\endcsname}%
\let\auto@bib@innerbib\@empty
\bibitem [{\citenamefont {Cambie}\ and\ \citenamefont
  {H{\"a}m{\"a}l{\"a}inen}(2025)}]{cambie2025packing}%
  \BibitemOpen
  \bibfield  {author} {\bibinfo {author} {\bibfnamefont {S.}~\bibnamefont
  {Cambie}}\ and\ \bibinfo {author} {\bibfnamefont {R.}~\bibnamefont
  {H{\"a}m{\"a}l{\"a}inen}},\ }\bibfield  {title} {\bibinfo {title} {Packing
  colourings in complete bipartite graphs and the inverse problem for
  correspondence packing},\ }\href@noop {} {\bibfield  {journal} {\bibinfo
  {journal} {Journal of Graph Theory}\ }\textbf {\bibinfo {volume} {109}},\
  \bibinfo {pages} {52} (\bibinfo {year} {2025})}\BibitemShut {NoStop}%
\bibitem [{\citenamefont {Bara}\ \emph {et~al.}(2025)\citenamefont {Bara},
  \citenamefont {Rasid},\ and\ \citenamefont {Artes~Jr}}]{bara2025independent}%
  \BibitemOpen
  \bibfield  {author} {\bibinfo {author} {\bibfnamefont {Z.~J.}\ \bibnamefont
  {Bara}}, \bibinfo {author} {\bibfnamefont {R.~A.}\ \bibnamefont {Rasid}},\
  and\ \bibinfo {author} {\bibfnamefont {R.~G.}\ \bibnamefont {Artes~Jr}},\
  }\bibfield  {title} {\bibinfo {title} {Independent vertex neighborhood
  polynomials: Behavior of graphs and some applications.},\ }\href@noop {}
  {\bibfield  {journal} {\bibinfo  {journal} {Advances \& Applications in
  Discrete Mathematics}\ }\textbf {\bibinfo {volume} {42}},\ \bibinfo {pages}
  {47} (\bibinfo {year} {2025})}\BibitemShut {NoStop}%
\bibitem [{\citenamefont {Punnen}\ and\ \citenamefont
  {Wang}(2016)}]{punnen2016bipartite}%
  \BibitemOpen
  \bibfield  {author} {\bibinfo {author} {\bibfnamefont {A.~P.}\ \bibnamefont
  {Punnen}}\ and\ \bibinfo {author} {\bibfnamefont {Y.}~\bibnamefont {Wang}},\
  }\bibfield  {title} {\bibinfo {title} {The bipartite quadratic assignment
  problem and extensions},\ }\href@noop {} {\bibfield  {journal} {\bibinfo
  {journal} {European Journal of Operational Research}\ }\textbf {\bibinfo
  {volume} {250}},\ \bibinfo {pages} {715} (\bibinfo {year}
  {2016})}\BibitemShut {NoStop}%
\bibitem [{\citenamefont {Ryan}\ \emph {et~al.}(2005)\citenamefont {Ryan},
  \citenamefont {Laforest}, \citenamefont {Boileau},\ and\ \citenamefont
  {Laflamme}}]{ryan2005experimental}%
  \BibitemOpen
  \bibfield  {author} {\bibinfo {author} {\bibfnamefont {C.~A.}\ \bibnamefont
  {Ryan}}, \bibinfo {author} {\bibfnamefont {M.}~\bibnamefont {Laforest}},
  \bibinfo {author} {\bibfnamefont {J.-C.}\ \bibnamefont {Boileau}},\ and\
  \bibinfo {author} {\bibfnamefont {R.}~\bibnamefont {Laflamme}},\ }\bibfield
  {title} {\bibinfo {title} {Experimental implementation of a discrete-time
  quantum random walk on an {NMR} quantum-information processor},\ }\href@noop
  {} {\bibfield  {journal} {\bibinfo  {journal} {Physical Review A—Atomic,
  Molecular, and Optical Physics}\ }\textbf {\bibinfo {volume} {72}},\ \bibinfo
  {pages} {062317} (\bibinfo {year} {2005})}\BibitemShut {NoStop}%
\bibitem [{\citenamefont {Lu}\ \emph {et~al.}(2010)\citenamefont {Lu},
  \citenamefont {Zhu}, \citenamefont {Zou}, \citenamefont {Peng}, \citenamefont
  {Yu}, \citenamefont {Zhang}, \citenamefont {Chen},\ and\ \citenamefont
  {Du}}]{lu2010experimental}%
  \BibitemOpen
  \bibfield  {author} {\bibinfo {author} {\bibfnamefont {D.}~\bibnamefont
  {Lu}}, \bibinfo {author} {\bibfnamefont {J.}~\bibnamefont {Zhu}}, \bibinfo
  {author} {\bibfnamefont {P.}~\bibnamefont {Zou}}, \bibinfo {author}
  {\bibfnamefont {X.}~\bibnamefont {Peng}}, \bibinfo {author} {\bibfnamefont
  {Y.}~\bibnamefont {Yu}}, \bibinfo {author} {\bibfnamefont {S.}~\bibnamefont
  {Zhang}}, \bibinfo {author} {\bibfnamefont {Q.}~\bibnamefont {Chen}},\ and\
  \bibinfo {author} {\bibfnamefont {J.}~\bibnamefont {Du}},\ }\bibfield
  {title} {\bibinfo {title} {Experimental implementation of a quantum
  random-walk search algorithm using strongly dipolar coupled spins},\
  }\href@noop {} {\bibfield  {journal} {\bibinfo  {journal} {Physical Review
  A—Atomic, Molecular, and Optical Physics}\ }\textbf {\bibinfo {volume}
  {81}},\ \bibinfo {pages} {022308} (\bibinfo {year} {2010})}\BibitemShut
  {NoStop}%
\bibitem [{\citenamefont {Gr{\"a}fe}\ \emph {et~al.}(2016)\citenamefont
  {Gr{\"a}fe}, \citenamefont {Heilmann}, \citenamefont {Lebugle}, \citenamefont
  {Guzman-Silva}, \citenamefont {Perez-Leija},\ and\ \citenamefont
  {Szameit}}]{grafe2016integrated}%
  \BibitemOpen
  \bibfield  {author} {\bibinfo {author} {\bibfnamefont {M.}~\bibnamefont
  {Gr{\"a}fe}}, \bibinfo {author} {\bibfnamefont {R.}~\bibnamefont {Heilmann}},
  \bibinfo {author} {\bibfnamefont {M.}~\bibnamefont {Lebugle}}, \bibinfo
  {author} {\bibfnamefont {D.}~\bibnamefont {Guzman-Silva}}, \bibinfo {author}
  {\bibfnamefont {A.}~\bibnamefont {Perez-Leija}},\ and\ \bibinfo {author}
  {\bibfnamefont {A.}~\bibnamefont {Szameit}},\ }\bibfield  {title} {\bibinfo
  {title} {Integrated photonic quantum walks},\ }\href@noop {} {\bibfield
  {journal} {\bibinfo  {journal} {Journal of Optics}\ }\textbf {\bibinfo
  {volume} {18}},\ \bibinfo {pages} {103002} (\bibinfo {year}
  {2016})}\BibitemShut {NoStop}%
\bibitem [{\citenamefont {Neves}\ and\ \citenamefont
  {Puentes}(2018)}]{neves2018photonic}%
  \BibitemOpen
  \bibfield  {author} {\bibinfo {author} {\bibfnamefont {L.}~\bibnamefont
  {Neves}}\ and\ \bibinfo {author} {\bibfnamefont {G.}~\bibnamefont
  {Puentes}},\ }\bibfield  {title} {\bibinfo {title} {Photonic discrete-time
  quantum walks and applications},\ }\href@noop {} {\bibfield  {journal}
  {\bibinfo  {journal} {Entropy}\ }\textbf {\bibinfo {volume} {20}},\ \bibinfo
  {pages} {731} (\bibinfo {year} {2018})}\BibitemShut {NoStop}%
\bibitem [{\citenamefont {Yan}\ \emph {et~al.}(2019)\citenamefont {Yan},
  \citenamefont {Zhang}, \citenamefont {Gong}, \citenamefont {Wu},
  \citenamefont {Zheng}, \citenamefont {Li}, \citenamefont {Wang},
  \citenamefont {Liang}, \citenamefont {Lin}, \citenamefont {Xu} \emph
  {et~al.}}]{yan2019strongly}%
  \BibitemOpen
  \bibfield  {author} {\bibinfo {author} {\bibfnamefont {Z.}~\bibnamefont
  {Yan}}, \bibinfo {author} {\bibfnamefont {Y.-R.}\ \bibnamefont {Zhang}},
  \bibinfo {author} {\bibfnamefont {M.}~\bibnamefont {Gong}}, \bibinfo {author}
  {\bibfnamefont {Y.}~\bibnamefont {Wu}}, \bibinfo {author} {\bibfnamefont
  {Y.}~\bibnamefont {Zheng}}, \bibinfo {author} {\bibfnamefont
  {S.}~\bibnamefont {Li}}, \bibinfo {author} {\bibfnamefont {C.}~\bibnamefont
  {Wang}}, \bibinfo {author} {\bibfnamefont {F.}~\bibnamefont {Liang}},
  \bibinfo {author} {\bibfnamefont {J.}~\bibnamefont {Lin}}, \bibinfo {author}
  {\bibfnamefont {Y.}~\bibnamefont {Xu}}, \emph {et~al.},\ }\bibfield  {title}
  {\bibinfo {title} {Strongly correlated quantum walks with a 12-qubit
  superconducting processor},\ }\href@noop {} {\bibfield  {journal} {\bibinfo
  {journal} {Science}\ }\textbf {\bibinfo {volume} {364}},\ \bibinfo {pages}
  {753} (\bibinfo {year} {2019})}\BibitemShut {NoStop}%
\bibitem [{\citenamefont {Gong}\ \emph {et~al.}(2021)\citenamefont {Gong},
  \citenamefont {Wang}, \citenamefont {Zha}, \citenamefont {Chen},
  \citenamefont {Huang}, \citenamefont {Wu}, \citenamefont {Zhu}, \citenamefont
  {Zhao}, \citenamefont {Li}, \citenamefont {Guo} \emph
  {et~al.}}]{gong2021quantum}%
  \BibitemOpen
  \bibfield  {author} {\bibinfo {author} {\bibfnamefont {M.}~\bibnamefont
  {Gong}}, \bibinfo {author} {\bibfnamefont {S.}~\bibnamefont {Wang}}, \bibinfo
  {author} {\bibfnamefont {C.}~\bibnamefont {Zha}}, \bibinfo {author}
  {\bibfnamefont {M.-C.}\ \bibnamefont {Chen}}, \bibinfo {author}
  {\bibfnamefont {H.-L.}\ \bibnamefont {Huang}}, \bibinfo {author}
  {\bibfnamefont {Y.}~\bibnamefont {Wu}}, \bibinfo {author} {\bibfnamefont
  {Q.}~\bibnamefont {Zhu}}, \bibinfo {author} {\bibfnamefont {Y.}~\bibnamefont
  {Zhao}}, \bibinfo {author} {\bibfnamefont {S.}~\bibnamefont {Li}}, \bibinfo
  {author} {\bibfnamefont {S.}~\bibnamefont {Guo}}, \emph {et~al.},\ }\bibfield
   {title} {\bibinfo {title} {Quantum walks on a programmable two-dimensional
  62-qubit superconducting processor},\ }\href@noop {} {\bibfield  {journal}
  {\bibinfo  {journal} {Science}\ }\textbf {\bibinfo {volume} {372}},\ \bibinfo
  {pages} {948} (\bibinfo {year} {2021})}\BibitemShut {NoStop}%
\bibitem [{\citenamefont {Zhou}\ \emph {et~al.}(2025)\citenamefont {Zhou},
  \citenamefont {Xu}, \citenamefont {Su}, \citenamefont {Zhang}, \citenamefont
  {Yu}, \citenamefont {Ma}, \citenamefont {Zhang}, \citenamefont {Shi},
  \citenamefont {Zheng}, \citenamefont {Pan} \emph
  {et~al.}}]{zhou2025demonstration}%
  \BibitemOpen
  \bibfield  {author} {\bibinfo {author} {\bibfnamefont {K.}~\bibnamefont
  {Zhou}}, \bibinfo {author} {\bibfnamefont {J.-W.}\ \bibnamefont {Xu}},
  \bibinfo {author} {\bibfnamefont {Q.-P.}\ \bibnamefont {Su}}, \bibinfo
  {author} {\bibfnamefont {Y.}~\bibnamefont {Zhang}}, \bibinfo {author}
  {\bibfnamefont {X.-M.}\ \bibnamefont {Yu}}, \bibinfo {author} {\bibfnamefont
  {Z.}~\bibnamefont {Ma}}, \bibinfo {author} {\bibfnamefont {H.-Y.}\
  \bibnamefont {Zhang}}, \bibinfo {author} {\bibfnamefont {H.-Y.}\ \bibnamefont
  {Shi}}, \bibinfo {author} {\bibfnamefont {W.}~\bibnamefont {Zheng}}, \bibinfo
  {author} {\bibfnamefont {S.}~\bibnamefont {Pan}}, \emph {et~al.},\ }\bibfield
   {title} {\bibinfo {title} {Demonstration of discrete-time quantum walks and
  observation of topological edge states in a superconducting qutrit chain},\
  }\href@noop {} {\bibfield  {journal} {\bibinfo  {journal} {Physical Review
  Letters}\ }\textbf {\bibinfo {volume} {135}},\ \bibinfo {pages} {250601}
  (\bibinfo {year} {2025})}\BibitemShut {NoStop}%
\bibitem [{\citenamefont {Dubovitskii}\ \emph {et~al.}(2026)\citenamefont
  {Dubovitskii}, \citenamefont {Utro}, \citenamefont {Bose}, \citenamefont
  {Parida}, \citenamefont {Maniscalco},\ and\ \citenamefont
  {Filippov}}]{dubovitskii2026experimental}%
  \BibitemOpen
  \bibfield  {author} {\bibinfo {author} {\bibfnamefont {V.}~\bibnamefont
  {Dubovitskii}}, \bibinfo {author} {\bibfnamefont {F.}~\bibnamefont {Utro}},
  \bibinfo {author} {\bibfnamefont {A.}~\bibnamefont {Bose}}, \bibinfo {author}
  {\bibfnamefont {L.}~\bibnamefont {Parida}}, \bibinfo {author} {\bibfnamefont
  {S.}~\bibnamefont {Maniscalco}},\ and\ \bibinfo {author} {\bibfnamefont
  {S.~N.}\ \bibnamefont {Filippov}},\ }\bibfield  {title} {\bibinfo {title}
  {Experimental implementation of a discrete-time quantum walk on biological
  networks},\ }\href@noop {} {\bibfield  {journal} {\bibinfo  {journal} {arXiv
  preprint arXiv:2602.24053}\ } (\bibinfo {year} {2026})}\BibitemShut {NoStop}%
\bibitem [{\citenamefont {Razzoli}\ \emph {et~al.}(2024)\citenamefont
  {Razzoli}, \citenamefont {Cenedese}, \citenamefont {Bondani},\ and\
  \citenamefont {Benenti}}]{razzoli2024efficient}%
  \BibitemOpen
  \bibfield  {author} {\bibinfo {author} {\bibfnamefont {L.}~\bibnamefont
  {Razzoli}}, \bibinfo {author} {\bibfnamefont {G.}~\bibnamefont {Cenedese}},
  \bibinfo {author} {\bibfnamefont {M.}~\bibnamefont {Bondani}},\ and\ \bibinfo
  {author} {\bibfnamefont {G.}~\bibnamefont {Benenti}},\ }\bibfield  {title}
  {\bibinfo {title} {Efficient implementation of discrete-time quantum walks on
  quantum computers},\ }\href@noop {} {\bibfield  {journal} {\bibinfo
  {journal} {Entropy}\ }\textbf {\bibinfo {volume} {26}},\ \bibinfo {pages}
  {313} (\bibinfo {year} {2024})}\BibitemShut {NoStop}%
\bibitem [{\citenamefont {Shakeel}(2020)}]{shakeel2020efficient}%
  \BibitemOpen
  \bibfield  {author} {\bibinfo {author} {\bibfnamefont {A.}~\bibnamefont
  {Shakeel}},\ }\bibfield  {title} {\bibinfo {title} {Efficient and scalable
  quantum walk algorithms via the quantum {Fourier} transform},\ }\href@noop {}
  {\bibfield  {journal} {\bibinfo  {journal} {Quantum Information Processing}\
  }\textbf {\bibinfo {volume} {19}},\ \bibinfo {pages} {323} (\bibinfo {year}
  {2020})}\BibitemShut {NoStop}%
\bibitem [{\citenamefont {Georgopoulos}\ \emph {et~al.}(2021)\citenamefont
  {Georgopoulos}, \citenamefont {Emary},\ and\ \citenamefont
  {Zuliani}}]{georgopoulos2021comparison}%
  \BibitemOpen
  \bibfield  {author} {\bibinfo {author} {\bibfnamefont {K.}~\bibnamefont
  {Georgopoulos}}, \bibinfo {author} {\bibfnamefont {C.}~\bibnamefont
  {Emary}},\ and\ \bibinfo {author} {\bibfnamefont {P.}~\bibnamefont
  {Zuliani}},\ }\bibfield  {title} {\bibinfo {title} {Comparison of
  quantum-walk implementations on noisy intermediate-scale quantum computers},\
  }\href@noop {} {\bibfield  {journal} {\bibinfo  {journal} {Physical Review
  A}\ }\textbf {\bibinfo {volume} {103}},\ \bibinfo {pages} {022408} (\bibinfo
  {year} {2021})}\BibitemShut {NoStop}%
\bibitem [{\citenamefont {Wing-Bocanegra}\ and\ \citenamefont
  {Venegas-Andraca}(2023)}]{wing2023circuit}%
  \BibitemOpen
  \bibfield  {author} {\bibinfo {author} {\bibfnamefont {A.}~\bibnamefont
  {Wing-Bocanegra}}\ and\ \bibinfo {author} {\bibfnamefont {S.~E.}\
  \bibnamefont {Venegas-Andraca}},\ }\bibfield  {title} {\bibinfo {title}
  {Circuit implementation of discrete-time quantum walks via the shunt
  decomposition method},\ }\href@noop {} {\bibfield  {journal} {\bibinfo
  {journal} {Quantum Information Processing}\ }\textbf {\bibinfo {volume}
  {22}},\ \bibinfo {pages} {146} (\bibinfo {year} {2023})}\BibitemShut
  {NoStop}%
\bibitem [{\citenamefont {Wadhia}\ \emph {et~al.}(2024)\citenamefont {Wadhia},
  \citenamefont {Chancellor},\ and\ \citenamefont {Kendon}}]{wadhia2024cycle}%
  \BibitemOpen
  \bibfield  {author} {\bibinfo {author} {\bibfnamefont {V.}~\bibnamefont
  {Wadhia}}, \bibinfo {author} {\bibfnamefont {N.}~\bibnamefont {Chancellor}},\
  and\ \bibinfo {author} {\bibfnamefont {V.}~\bibnamefont {Kendon}},\
  }\bibfield  {title} {\bibinfo {title} {Cycle discrete-time quantum walks on a
  noisy quantum computer},\ }\href@noop {} {\bibfield  {journal} {\bibinfo
  {journal} {The European Physical Journal D}\ }\textbf {\bibinfo {volume}
  {78}},\ \bibinfo {pages} {29} (\bibinfo {year} {2024})}\BibitemShut {NoStop}%
\bibitem [{\citenamefont {Aharonov}\ \emph {et~al.}(2001)\citenamefont
  {Aharonov}, \citenamefont {Ambainis}, \citenamefont {Kempe},\ and\
  \citenamefont {Vazirani}}]{Aharonov2001QuantumWalks}%
  \BibitemOpen
  \bibfield  {author} {\bibinfo {author} {\bibfnamefont {D.}~\bibnamefont
  {Aharonov}}, \bibinfo {author} {\bibfnamefont {A.}~\bibnamefont {Ambainis}},
  \bibinfo {author} {\bibfnamefont {J.}~\bibnamefont {Kempe}},\ and\ \bibinfo
  {author} {\bibfnamefont {U.}~\bibnamefont {Vazirani}},\ }\bibfield  {title}
  {\bibinfo {title} {Quantum walks on graphs},\ }in\ \href@noop {} {\emph
  {\bibinfo {booktitle} {Proceedings of the Thirty-Third Annual ACM Symposium
  on Theory of Computing}}}\ (\bibinfo  {publisher} {ACM},\ \bibinfo {address}
  {New York, NY, USA},\ \bibinfo {year} {2001})\ pp.\ \bibinfo {pages}
  {50--59}\BibitemShut {NoStop}%
\bibitem [{\citenamefont {Childs}\ \emph {et~al.}(2003)\citenamefont {Childs},
  \citenamefont {Cleve}, \citenamefont {Deotto}, \citenamefont {Farhi},
  \citenamefont {Gutmann},\ and\ \citenamefont
  {Spielman}}]{childs2003exponential}%
  \BibitemOpen
  \bibfield  {author} {\bibinfo {author} {\bibfnamefont {A.~M.}\ \bibnamefont
  {Childs}}, \bibinfo {author} {\bibfnamefont {R.}~\bibnamefont {Cleve}},
  \bibinfo {author} {\bibfnamefont {E.}~\bibnamefont {Deotto}}, \bibinfo
  {author} {\bibfnamefont {E.}~\bibnamefont {Farhi}}, \bibinfo {author}
  {\bibfnamefont {S.}~\bibnamefont {Gutmann}},\ and\ \bibinfo {author}
  {\bibfnamefont {D.~A.}\ \bibnamefont {Spielman}},\ }\bibfield  {title}
  {\bibinfo {title} {Exponential algorithmic speedup by a quantum walk},\ }in\
  \href@noop {} {\emph {\bibinfo {booktitle} {Proceedings of the thirty-fifth
  annual ACM symposium on Theory of computing}}}\ (\bibinfo {year} {2003})\
  pp.\ \bibinfo {pages} {59--68}\BibitemShut {NoStop}%
\bibitem [{\citenamefont {Childs}\ and\ \citenamefont
  {Goldstone}(2004)}]{Childs2004SpatialSearch}%
  \BibitemOpen
  \bibfield  {author} {\bibinfo {author} {\bibfnamefont {A.~M.}\ \bibnamefont
  {Childs}}\ and\ \bibinfo {author} {\bibfnamefont {J.}~\bibnamefont
  {Goldstone}},\ }\bibfield  {title} {\bibinfo {title} {Spatial search by
  quantum walk},\ }\href {https://doi.org/10.1103/PhysRevA.70.022314}
  {\bibfield  {journal} {\bibinfo  {journal} {Physical Review A}\ }\textbf
  {\bibinfo {volume} {70}},\ \bibinfo {pages} {022314} (\bibinfo {year}
  {2004})}\BibitemShut {NoStop}%
\bibitem [{\citenamefont {Chakraborty}\ \emph {et~al.}(2016)\citenamefont
  {Chakraborty}, \citenamefont {Novo}, \citenamefont {Ambainis},\ and\
  \citenamefont {Omar}}]{Chakraborty2016OptimalSpatialSearch}%
  \BibitemOpen
  \bibfield  {author} {\bibinfo {author} {\bibfnamefont {S.}~\bibnamefont
  {Chakraborty}}, \bibinfo {author} {\bibfnamefont {L.}~\bibnamefont {Novo}},
  \bibinfo {author} {\bibfnamefont {A.}~\bibnamefont {Ambainis}},\ and\
  \bibinfo {author} {\bibfnamefont {Y.}~\bibnamefont {Omar}},\ }\bibfield
  {title} {\bibinfo {title} {Spatial search by quantum walk is optimal for
  almost all graphs},\ }\href {https://doi.org/10.1103/PhysRevLett.116.100501}
  {\bibfield  {journal} {\bibinfo  {journal} {Physical Review Letters}\
  }\textbf {\bibinfo {volume} {116}},\ \bibinfo {pages} {100501} (\bibinfo
  {year} {2016})}\BibitemShut {NoStop}%
\bibitem [{\citenamefont {Ambainis}\ \emph {et~al.}(2020)\citenamefont
  {Ambainis}, \citenamefont {Gily{\'e}n}, \citenamefont {Jeffery},\ and\
  \citenamefont {Kokiānis}}]{Ambainis2020MarkedVertices}%
  \BibitemOpen
  \bibfield  {author} {\bibinfo {author} {\bibfnamefont {A.}~\bibnamefont
  {Ambainis}}, \bibinfo {author} {\bibfnamefont {A.}~\bibnamefont
  {Gily{\'e}n}}, \bibinfo {author} {\bibfnamefont {S.}~\bibnamefont
  {Jeffery}},\ and\ \bibinfo {author} {\bibfnamefont {D.}~\bibnamefont
  {Kokiānis}},\ }\bibfield  {title} {\bibinfo {title} {Quadratic speedup for
  finding marked vertices by quantum walks},\ }in\ \href@noop {} {\emph
  {\bibinfo {booktitle} {Proceedings of the 52nd Annual ACM SIGACT Symposium on
  Theory of Computing}}}\ (\bibinfo  {publisher} {ACM},\ \bibinfo {address}
  {New York, NY, USA},\ \bibinfo {year} {2020})\ pp.\ \bibinfo {pages}
  {1119--1132}\BibitemShut {NoStop}%
\bibitem [{\citenamefont {Apers}\ \emph {et~al.}(2022)\citenamefont {Apers},
  \citenamefont {Chakraborty}, \citenamefont {Novo},\ and\ \citenamefont
  {Roland}}]{apers2022quadratic}%
  \BibitemOpen
  \bibfield  {author} {\bibinfo {author} {\bibfnamefont {S.}~\bibnamefont
  {Apers}}, \bibinfo {author} {\bibfnamefont {S.}~\bibnamefont {Chakraborty}},
  \bibinfo {author} {\bibfnamefont {L.}~\bibnamefont {Novo}},\ and\ \bibinfo
  {author} {\bibfnamefont {J.}~\bibnamefont {Roland}},\ }\bibfield  {title}
  {\bibinfo {title} {Quadratic speedup for spatial search by continuous-time
  quantum walk},\ }\href@noop {} {\bibfield  {journal} {\bibinfo  {journal}
  {Physical Review Letters}\ }\textbf {\bibinfo {volume} {129}},\ \bibinfo
  {pages} {160502} (\bibinfo {year} {2022})}\BibitemShut {NoStop}%
\bibitem [{\citenamefont {Peng}\ \emph
  {et~al.}(2024{\natexlab{a}})\citenamefont {Peng}, \citenamefont {Li},\ and\
  \citenamefont {Sun}}]{peng2024search}%
  \BibitemOpen
  \bibfield  {author} {\bibinfo {author} {\bibfnamefont {F.}~\bibnamefont
  {Peng}}, \bibinfo {author} {\bibfnamefont {M.}~\bibnamefont {Li}},\ and\
  \bibinfo {author} {\bibfnamefont {X.}~\bibnamefont {Sun}},\ }\bibfield
  {title} {\bibinfo {title} {Search algorithm on strongly regular graph by
  lackadaisical quantum walk},\ }\href@noop {} {\bibfield  {journal} {\bibinfo
  {journal} {Journal of Physics A: Mathematical and Theoretical}\ }\textbf
  {\bibinfo {volume} {57}},\ \bibinfo {pages} {135301} (\bibinfo {year}
  {2024}{\natexlab{a}})}\BibitemShut {NoStop}%
\bibitem [{\citenamefont {Peng}\ \emph
  {et~al.}(2024{\natexlab{b}})\citenamefont {Peng}, \citenamefont {Li},\ and\
  \citenamefont {Sun}}]{peng2024lackadaisical}%
  \BibitemOpen
  \bibfield  {author} {\bibinfo {author} {\bibfnamefont {F.}~\bibnamefont
  {Peng}}, \bibinfo {author} {\bibfnamefont {M.}~\bibnamefont {Li}},\ and\
  \bibinfo {author} {\bibfnamefont {X.}~\bibnamefont {Sun}},\ }\bibfield
  {title} {\bibinfo {title} {Lackadaisical discrete-time quantum walk on
  {Johnson} graph},\ }\href@noop {} {\bibfield  {journal} {\bibinfo  {journal}
  {Physica A: Statistical Mechanics and its Applications}\ }\textbf {\bibinfo
  {volume} {635}},\ \bibinfo {pages} {129495} (\bibinfo {year}
  {2024}{\natexlab{b}})}\BibitemShut {NoStop}%
\bibitem [{\citenamefont {H{\o}yer}(2000)}]{Hoyer2000ArbitraryPhases}%
  \BibitemOpen
  \bibfield  {author} {\bibinfo {author} {\bibfnamefont {P.}~\bibnamefont
  {H{\o}yer}},\ }\bibfield  {title} {\bibinfo {title} {Arbitrary phases in
  quantum amplitude amplification},\ }\href
  {https://doi.org/10.1103/PhysRevA.62.052304} {\bibfield  {journal} {\bibinfo
  {journal} {Physical Review A}\ }\textbf {\bibinfo {volume} {62}},\ \bibinfo
  {pages} {052304} (\bibinfo {year} {2000})}\BibitemShut {NoStop}%
\bibitem [{\citenamefont {Long}(2001)}]{Long2001ZeroFailureGrover}%
  \BibitemOpen
  \bibfield  {author} {\bibinfo {author} {\bibfnamefont {G.-L.}\ \bibnamefont
  {Long}},\ }\bibfield  {title} {\bibinfo {title} {Grover algorithm with zero
  theoretical failure rate},\ }\href
  {https://doi.org/10.1103/PhysRevA.64.022307} {\bibfield  {journal} {\bibinfo
  {journal} {Physical Review A}\ }\textbf {\bibinfo {volume} {64}},\ \bibinfo
  {pages} {022307} (\bibinfo {year} {2001})}\BibitemShut {NoStop}%
\bibitem [{\citenamefont {Marsh}\ and\ \citenamefont
  {Wang}(2021)}]{Marsh2021DeterministicSpatialSearch}%
  \BibitemOpen
  \bibfield  {author} {\bibinfo {author} {\bibfnamefont {S.}~\bibnamefont
  {Marsh}}\ and\ \bibinfo {author} {\bibfnamefont {J.~B.}\ \bibnamefont
  {Wang}},\ }\bibfield  {title} {\bibinfo {title} {Deterministic spatial search
  using alternating quantum walks},\ }\href
  {https://doi.org/10.1103/PhysRevA.104.022216} {\bibfield  {journal} {\bibinfo
   {journal} {Physical Review A}\ }\textbf {\bibinfo {volume} {104}},\ \bibinfo
  {pages} {022216} (\bibinfo {year} {2021})}\BibitemShut {NoStop}%
\bibitem [{\citenamefont {Qu}\ \emph {et~al.}(2022)\citenamefont {Qu},
  \citenamefont {Marsh}, \citenamefont {Wang}, \citenamefont {Xiao},
  \citenamefont {Wang},\ and\ \citenamefont
  {Xue}}]{Qu2022DeterministicStarGraphs}%
  \BibitemOpen
  \bibfield  {author} {\bibinfo {author} {\bibfnamefont {D.}~\bibnamefont
  {Qu}}, \bibinfo {author} {\bibfnamefont {S.}~\bibnamefont {Marsh}}, \bibinfo
  {author} {\bibfnamefont {K.}~\bibnamefont {Wang}}, \bibinfo {author}
  {\bibfnamefont {L.}~\bibnamefont {Xiao}}, \bibinfo {author} {\bibfnamefont
  {J.}~\bibnamefont {Wang}},\ and\ \bibinfo {author} {\bibfnamefont
  {P.}~\bibnamefont {Xue}},\ }\bibfield  {title} {\bibinfo {title}
  {Deterministic search on star graphs via quantum walks},\ }\href
  {https://doi.org/10.1103/PhysRevLett.128.050501} {\bibfield  {journal}
  {\bibinfo  {journal} {Physical Review Letters}\ }\textbf {\bibinfo {volume}
  {128}},\ \bibinfo {pages} {050501} (\bibinfo {year} {2022})}\BibitemShut
  {NoStop}%
\bibitem [{\citenamefont {Wang}\ \emph {et~al.}(2025)\citenamefont {Wang},
  \citenamefont {Jiang}, \citenamefont {Feng},\ and\ \citenamefont
  {Li}}]{wang2025unifying}%
  \BibitemOpen
  \bibfield  {author} {\bibinfo {author} {\bibfnamefont {Q.}~\bibnamefont
  {Wang}}, \bibinfo {author} {\bibfnamefont {Y.}~\bibnamefont {Jiang}},
  \bibinfo {author} {\bibfnamefont {S.}~\bibnamefont {Feng}},\ and\ \bibinfo
  {author} {\bibfnamefont {L.}~\bibnamefont {Li}},\ }\bibfield  {title}
  {\bibinfo {title} {Unifying quantum spatial search, state transfer, and
  uniform sampling on graphs},\ }\href@noop {} {\bibfield  {journal} {\bibinfo
  {journal} {Physical Review A}\ }\textbf {\bibinfo {volume} {111}},\ \bibinfo
  {pages} {042608} (\bibinfo {year} {2025})}\BibitemShut {NoStop}%
\bibitem [{\citenamefont {Peng}\ \emph
  {et~al.}(2024{\natexlab{c}})\citenamefont {Peng}, \citenamefont {Li},\ and\
  \citenamefont {Sun}}]{Peng2024DeterministicDTQW}%
  \BibitemOpen
  \bibfield  {author} {\bibinfo {author} {\bibfnamefont {F.}~\bibnamefont
  {Peng}}, \bibinfo {author} {\bibfnamefont {M.}~\bibnamefont {Li}},\ and\
  \bibinfo {author} {\bibfnamefont {X.}~\bibnamefont {Sun}},\ }\bibfield
  {title} {\bibinfo {title} {Deterministic discrete-time quantum walk search on
  complete bipartite graphs},\ }\href
  {https://doi.org/10.1103/PhysRevResearch.6.033042} {\bibfield  {journal}
  {\bibinfo  {journal} {Physical Review Research}\ }\textbf {\bibinfo {volume}
  {6}},\ \bibinfo {pages} {033042} (\bibinfo {year}
  {2024}{\natexlab{c}})}\BibitemShut {NoStop}%
\bibitem [{\citenamefont {{Jiang}}\ \emph {et~al.}(2026)\citenamefont
  {{Jiang}}, \citenamefont {{Cai}}, \citenamefont {{Huang}}, \citenamefont
  {{Zhou}}, \citenamefont {{Zhang}}, \citenamefont {{Bei}}, \citenamefont
  {{Cai}}, \citenamefont {{Cao}}, \citenamefont {{Chen}}, \citenamefont
  {{Chen}}, \citenamefont {{Chen}}, \citenamefont {{Chen}}, \citenamefont
  {{Chen}}, \citenamefont {{Chen}}, \citenamefont {{Chen}}, \citenamefont
  {{Chen}}, \citenamefont {{Chu}}, \citenamefont {{Deng}}, \citenamefont
  {{Deng}}, \citenamefont {{Ding}}, \citenamefont {{Ding}}, \citenamefont
  {{Ding}}, \citenamefont {{Dong}}, \citenamefont {{Fan}}, \citenamefont
  {{Fan}}, \citenamefont {{Fu}}, \citenamefont {{Gao}}, \citenamefont {{Ge}},
  \citenamefont {{Gui}}, \citenamefont {{Guo}}, \citenamefont {{Guo}},
  \citenamefont {{Guo}}, \citenamefont {{Han}}, \citenamefont {{He}},
  \citenamefont {{Hong}}, \citenamefont {{Hu}}, \citenamefont {{Huang}},
  \citenamefont {{Huo}}, \citenamefont {{Jiang}}, \citenamefont {{Jin}},
  \citenamefont {{Leng}}, \citenamefont {{Li}}, \citenamefont {{Li}},
  \citenamefont {{Li}}, \citenamefont {{Li}}, \citenamefont {{Li}},
  \citenamefont {{Li}}, \citenamefont {{Li}}, \citenamefont {{Li}},
  \citenamefont {{Li}}, \citenamefont {{Li}}, \citenamefont {{Li}},
  \citenamefont {{Li}}, \citenamefont {{Liang}}, \citenamefont {{Liang}},
  \citenamefont {{Liao}}, \citenamefont {{Lin}}, \citenamefont {{Lin}},
  \citenamefont {{Liu}}, \citenamefont {{Liu}}, \citenamefont {{Liu}},
  \citenamefont {{Liu}}, \citenamefont {{Liu}}, \citenamefont {{Liu}},
  \citenamefont {{Lou}}, \citenamefont {{Ma}}, \citenamefont {{Meng}},
  \citenamefont {{Mou}}, \citenamefont {{Nan}}, \citenamefont {{Nie}},
  \citenamefont {{Nie}}, \citenamefont {{Ning}}, \citenamefont {{Niu}},
  \citenamefont {{Peng}}, \citenamefont {{Qian}}, \citenamefont {{Rong}},
  \citenamefont {{Rong}}, \citenamefont {{Shen}}, \citenamefont {{Shen}},
  \citenamefont {{Su}}, \citenamefont {{Su}}, \citenamefont {{Sun}},
  \citenamefont {{Sun}}, \citenamefont {{Sun}}, \citenamefont {{Sun}},
  \citenamefont {{Tan}}, \citenamefont {{Tan}}, \citenamefont {{Tang}},
  \citenamefont {{Tu}}, \citenamefont {{Wang}}, \citenamefont {{Wang}},
  \citenamefont {{Wang}}, \citenamefont {{Wang}}, \citenamefont {{Wang}},
  \citenamefont {{Wang}}, \citenamefont {{Wang}}, \citenamefont {{Wang}},
  \citenamefont {{Wang}}, \citenamefont {{Wang}}, \citenamefont {{Wang}},
  \citenamefont {{Wang}}, \citenamefont {{Wang}}, \citenamefont {{Wei}},
  \citenamefont {{Wei}}, \citenamefont {{Wu}}, \citenamefont {{Wu}},
  \citenamefont {{Wu}}, \citenamefont {{Wu}}, \citenamefont {{Xie}},
  \citenamefont {{Xin}}, \citenamefont {{Xu}}, \citenamefont {{Xue}},
  \citenamefont {{Yan}}, \citenamefont {{Yang}}, \citenamefont {{Yang}},
  \citenamefont {{Yang}}, \citenamefont {{Ye}}, \citenamefont {{Ye}},
  \citenamefont {{Ying}}, \citenamefont {{Yu}}, \citenamefont {{Yu}},
  \citenamefont {{Yu}}, \citenamefont {{Zeng}}, \citenamefont {{Zha}},
  \citenamefont {{Zhan}}, \citenamefont {{Zhang}}, \citenamefont {{Zhang}},
  \citenamefont {{Zhang}}, \citenamefont {{Zhang}}, \citenamefont {{Zhang}},
  \citenamefont {{Zhang}}, \citenamefont {{Zhang}}, \citenamefont {{Zhao}},
  \citenamefont {{Zhao}}, \citenamefont {{Zhao}}, \citenamefont {{Zhao}},
  \citenamefont {{Zhao}}, \citenamefont {{Zheng}}, \citenamefont {{Zhou}},
  \citenamefont {{Zhou}}, \citenamefont {{Zhou}}, \citenamefont {{Zhou}},
  \citenamefont {{Zhou}}, \citenamefont {{Zhou}}, \citenamefont {{Zhu}},
  \citenamefont {{Zhu}}, \citenamefont {{Zou}}, \citenamefont {{Zou}},
  \citenamefont {{Zhang}}, \citenamefont {{Lu}}, \citenamefont {{Peng}},
  \citenamefont {{Yuan}}, \citenamefont {{Gong}}, \citenamefont {{Zhu}},\ and\
  \citenamefont {{Pan}}}]{jiang2025generation95qubitgenuineentanglement}%
  \BibitemOpen
  \bibfield  {author} {\bibinfo {author} {\bibfnamefont {T.}~\bibnamefont
  {{Jiang}}}, \bibinfo {author} {\bibfnamefont {J.}~\bibnamefont {{Cai}}},
  \bibinfo {author} {\bibfnamefont {J.}~\bibnamefont {{Huang}}}, \bibinfo
  {author} {\bibfnamefont {N.}~\bibnamefont {{Zhou}}}, \bibinfo {author}
  {\bibfnamefont {Y.}~\bibnamefont {{Zhang}}}, \bibinfo {author} {\bibfnamefont
  {J.}~\bibnamefont {{Bei}}}, \bibinfo {author} {\bibfnamefont
  {G.}~\bibnamefont {{Cai}}}, \bibinfo {author} {\bibfnamefont
  {S.}~\bibnamefont {{Cao}}}, \bibinfo {author} {\bibfnamefont
  {F.}~\bibnamefont {{Chen}}}, \bibinfo {author} {\bibfnamefont
  {J.}~\bibnamefont {{Chen}}}, \bibinfo {author} {\bibfnamefont
  {K.}~\bibnamefont {{Chen}}}, \bibinfo {author} {\bibfnamefont
  {X.}~\bibnamefont {{Chen}}}, \bibinfo {author} {\bibfnamefont
  {X.}~\bibnamefont {{Chen}}}, \bibinfo {author} {\bibfnamefont
  {Z.}~\bibnamefont {{Chen}}}, \bibinfo {author} {\bibfnamefont
  {Z.}~\bibnamefont {{Chen}}}, \bibinfo {author} {\bibfnamefont
  {Z.}~\bibnamefont {{Chen}}}, \bibinfo {author} {\bibfnamefont
  {W.}~\bibnamefont {{Chu}}}, \bibinfo {author} {\bibfnamefont
  {H.}~\bibnamefont {{Deng}}}, \bibinfo {author} {\bibfnamefont
  {Z.}~\bibnamefont {{Deng}}}, \bibinfo {author} {\bibfnamefont
  {P.}~\bibnamefont {{Ding}}}, \bibinfo {author} {\bibfnamefont
  {X.}~\bibnamefont {{Ding}}}, \bibinfo {author} {\bibfnamefont
  {Z.}~\bibnamefont {{Ding}}}, \bibinfo {author} {\bibfnamefont
  {S.}~\bibnamefont {{Dong}}}, \bibinfo {author} {\bibfnamefont
  {B.}~\bibnamefont {{Fan}}}, \bibinfo {author} {\bibfnamefont
  {D.}~\bibnamefont {{Fan}}}, \bibinfo {author} {\bibfnamefont
  {Y.}~\bibnamefont {{Fu}}}, \bibinfo {author} {\bibfnamefont {D.}~\bibnamefont
  {{Gao}}}, \bibinfo {author} {\bibfnamefont {L.}~\bibnamefont {{Ge}}},
  \bibinfo {author} {\bibfnamefont {J.}~\bibnamefont {{Gui}}}, \bibinfo
  {author} {\bibfnamefont {C.}~\bibnamefont {{Guo}}}, \bibinfo {author}
  {\bibfnamefont {S.}~\bibnamefont {{Guo}}}, \bibinfo {author} {\bibfnamefont
  {X.}~\bibnamefont {{Guo}}}, \bibinfo {author} {\bibfnamefont
  {L.}~\bibnamefont {{Han}}}, \bibinfo {author} {\bibfnamefont
  {T.}~\bibnamefont {{He}}}, \bibinfo {author} {\bibfnamefont {L.}~\bibnamefont
  {{Hong}}}, \bibinfo {author} {\bibfnamefont {Y.}~\bibnamefont {{Hu}}},
  \bibinfo {author} {\bibfnamefont {H.-L.}\ \bibnamefont {{Huang}}}, \bibinfo
  {author} {\bibfnamefont {Y.-H.}\ \bibnamefont {{Huo}}}, \bibinfo {author}
  {\bibfnamefont {Z.}~\bibnamefont {{Jiang}}}, \bibinfo {author} {\bibfnamefont
  {H.}~\bibnamefont {{Jin}}}, \bibinfo {author} {\bibfnamefont
  {Y.}~\bibnamefont {{Leng}}}, \bibinfo {author} {\bibfnamefont
  {D.}~\bibnamefont {{Li}}}, \bibinfo {author} {\bibfnamefont {D.}~\bibnamefont
  {{Li}}}, \bibinfo {author} {\bibfnamefont {F.}~\bibnamefont {{Li}}}, \bibinfo
  {author} {\bibfnamefont {J.}~\bibnamefont {{Li}}}, \bibinfo {author}
  {\bibfnamefont {J.}~\bibnamefont {{Li}}}, \bibinfo {author} {\bibfnamefont
  {J.}~\bibnamefont {{Li}}}, \bibinfo {author} {\bibfnamefont {J.}~\bibnamefont
  {{Li}}}, \bibinfo {author} {\bibfnamefont {N.}~\bibnamefont {{Li}}}, \bibinfo
  {author} {\bibfnamefont {S.}~\bibnamefont {{Li}}}, \bibinfo {author}
  {\bibfnamefont {W.}~\bibnamefont {{Li}}}, \bibinfo {author} {\bibfnamefont
  {Y.}~\bibnamefont {{Li}}}, \bibinfo {author} {\bibfnamefont {Y.}~\bibnamefont
  {{Li}}}, \bibinfo {author} {\bibfnamefont {F.}~\bibnamefont {{Liang}}},
  \bibinfo {author} {\bibfnamefont {X.}~\bibnamefont {{Liang}}}, \bibinfo
  {author} {\bibfnamefont {N.}~\bibnamefont {{Liao}}}, \bibinfo {author}
  {\bibfnamefont {J.}~\bibnamefont {{Lin}}}, \bibinfo {author} {\bibfnamefont
  {W.}~\bibnamefont {{Lin}}}, \bibinfo {author} {\bibfnamefont
  {D.}~\bibnamefont {{Liu}}}, \bibinfo {author} {\bibfnamefont
  {H.}~\bibnamefont {{Liu}}}, \bibinfo {author} {\bibfnamefont
  {M.}~\bibnamefont {{Liu}}}, \bibinfo {author} {\bibfnamefont
  {X.}~\bibnamefont {{Liu}}}, \bibinfo {author} {\bibfnamefont
  {X.}~\bibnamefont {{Liu}}}, \bibinfo {author} {\bibfnamefont
  {Y.}~\bibnamefont {{Liu}}}, \bibinfo {author} {\bibfnamefont
  {H.}~\bibnamefont {{Lou}}}, \bibinfo {author} {\bibfnamefont
  {Y.}~\bibnamefont {{Ma}}}, \bibinfo {author} {\bibfnamefont {L.}~\bibnamefont
  {{Meng}}}, \bibinfo {author} {\bibfnamefont {H.}~\bibnamefont {{Mou}}},
  \bibinfo {author} {\bibfnamefont {K.}~\bibnamefont {{Nan}}}, \bibinfo
  {author} {\bibfnamefont {B.}~\bibnamefont {{Nie}}}, \bibinfo {author}
  {\bibfnamefont {M.}~\bibnamefont {{Nie}}}, \bibinfo {author} {\bibfnamefont
  {J.}~\bibnamefont {{Ning}}}, \bibinfo {author} {\bibfnamefont
  {L.}~\bibnamefont {{Niu}}}, \bibinfo {author} {\bibfnamefont
  {W.}~\bibnamefont {{Peng}}}, \bibinfo {author} {\bibfnamefont
  {H.}~\bibnamefont {{Qian}}}, \bibinfo {author} {\bibfnamefont
  {H.}~\bibnamefont {{Rong}}}, \bibinfo {author} {\bibfnamefont
  {T.}~\bibnamefont {{Rong}}}, \bibinfo {author} {\bibfnamefont
  {H.}~\bibnamefont {{Shen}}}, \bibinfo {author} {\bibfnamefont
  {Q.}~\bibnamefont {{Shen}}}, \bibinfo {author} {\bibfnamefont
  {H.}~\bibnamefont {{Su}}}, \bibinfo {author} {\bibfnamefont {F.}~\bibnamefont
  {{Su}}}, \bibinfo {author} {\bibfnamefont {C.}~\bibnamefont {{Sun}}},
  \bibinfo {author} {\bibfnamefont {L.}~\bibnamefont {{Sun}}}, \bibinfo
  {author} {\bibfnamefont {T.}~\bibnamefont {{Sun}}}, \bibinfo {author}
  {\bibfnamefont {Y.}~\bibnamefont {{Sun}}}, \bibinfo {author} {\bibfnamefont
  {Y.}~\bibnamefont {{Tan}}}, \bibinfo {author} {\bibfnamefont
  {J.}~\bibnamefont {{Tan}}}, \bibinfo {author} {\bibfnamefont
  {L.}~\bibnamefont {{Tang}}}, \bibinfo {author} {\bibfnamefont
  {W.}~\bibnamefont {{Tu}}}, \bibinfo {author} {\bibfnamefont {J.}~\bibnamefont
  {{Wang}}}, \bibinfo {author} {\bibfnamefont {B.}~\bibnamefont {{Wang}}},
  \bibinfo {author} {\bibfnamefont {C.}~\bibnamefont {{Wang}}}, \bibinfo
  {author} {\bibfnamefont {C.}~\bibnamefont {{Wang}}}, \bibinfo {author}
  {\bibfnamefont {C.}~\bibnamefont {{Wang}}}, \bibinfo {author} {\bibfnamefont
  {J.}~\bibnamefont {{Wang}}}, \bibinfo {author} {\bibfnamefont
  {L.}~\bibnamefont {{Wang}}}, \bibinfo {author} {\bibfnamefont
  {R.}~\bibnamefont {{Wang}}}, \bibinfo {author} {\bibfnamefont
  {S.}~\bibnamefont {{Wang}}}, \bibinfo {author} {\bibfnamefont
  {X.}~\bibnamefont {{Wang}}}, \bibinfo {author} {\bibfnamefont
  {X.}~\bibnamefont {{Wang}}}, \bibinfo {author} {\bibfnamefont
  {X.}~\bibnamefont {{Wang}}}, \bibinfo {author} {\bibfnamefont
  {Y.}~\bibnamefont {{Wang}}}, \bibinfo {author} {\bibfnamefont
  {Z.}~\bibnamefont {{Wei}}}, \bibinfo {author} {\bibfnamefont
  {J.}~\bibnamefont {{Wei}}}, \bibinfo {author} {\bibfnamefont
  {D.}~\bibnamefont {{Wu}}}, \bibinfo {author} {\bibfnamefont {G.}~\bibnamefont
  {{Wu}}}, \bibinfo {author} {\bibfnamefont {J.}~\bibnamefont {{Wu}}}, \bibinfo
  {author} {\bibfnamefont {Y.}~\bibnamefont {{Wu}}}, \bibinfo {author}
  {\bibfnamefont {S.}~\bibnamefont {{Xie}}}, \bibinfo {author} {\bibfnamefont
  {L.}~\bibnamefont {{Xin}}}, \bibinfo {author} {\bibfnamefont
  {Y.}~\bibnamefont {{Xu}}}, \bibinfo {author} {\bibfnamefont {C.}~\bibnamefont
  {{Xue}}}, \bibinfo {author} {\bibfnamefont {K.}~\bibnamefont {{Yan}}},
  \bibinfo {author} {\bibfnamefont {W.}~\bibnamefont {{Yang}}}, \bibinfo
  {author} {\bibfnamefont {X.}~\bibnamefont {{Yang}}}, \bibinfo {author}
  {\bibfnamefont {Y.}~\bibnamefont {{Yang}}}, \bibinfo {author} {\bibfnamefont
  {Y.}~\bibnamefont {{Ye}}}, \bibinfo {author} {\bibfnamefont {Z.}~\bibnamefont
  {{Ye}}}, \bibinfo {author} {\bibfnamefont {C.}~\bibnamefont {{Ying}}},
  \bibinfo {author} {\bibfnamefont {J.}~\bibnamefont {{Yu}}}, \bibinfo {author}
  {\bibfnamefont {Q.}~\bibnamefont {{Yu}}}, \bibinfo {author} {\bibfnamefont
  {W.}~\bibnamefont {{Yu}}}, \bibinfo {author} {\bibfnamefont {X.}~\bibnamefont
  {{Zeng}}}, \bibinfo {author} {\bibfnamefont {C.}~\bibnamefont {{Zha}}},
  \bibinfo {author} {\bibfnamefont {S.}~\bibnamefont {{Zhan}}}, \bibinfo
  {author} {\bibfnamefont {F.}~\bibnamefont {{Zhang}}}, \bibinfo {author}
  {\bibfnamefont {H.}~\bibnamefont {{Zhang}}}, \bibinfo {author} {\bibfnamefont
  {K.}~\bibnamefont {{Zhang}}}, \bibinfo {author} {\bibfnamefont
  {W.}~\bibnamefont {{Zhang}}}, \bibinfo {author} {\bibfnamefont
  {Y.}~\bibnamefont {{Zhang}}}, \bibinfo {author} {\bibfnamefont
  {Y.}~\bibnamefont {{Zhang}}}, \bibinfo {author} {\bibfnamefont
  {L.}~\bibnamefont {{Zhang}}}, \bibinfo {author} {\bibfnamefont
  {G.}~\bibnamefont {{Zhao}}}, \bibinfo {author} {\bibfnamefont
  {P.}~\bibnamefont {{Zhao}}}, \bibinfo {author} {\bibfnamefont
  {X.}~\bibnamefont {{Zhao}}}, \bibinfo {author} {\bibfnamefont
  {Y.}~\bibnamefont {{Zhao}}}, \bibinfo {author} {\bibfnamefont
  {Z.}~\bibnamefont {{Zhao}}}, \bibinfo {author} {\bibfnamefont
  {L.}~\bibnamefont {{Zheng}}}, \bibinfo {author} {\bibfnamefont
  {F.}~\bibnamefont {{Zhou}}}, \bibinfo {author} {\bibfnamefont
  {L.}~\bibnamefont {{Zhou}}}, \bibinfo {author} {\bibfnamefont
  {N.}~\bibnamefont {{Zhou}}}, \bibinfo {author} {\bibfnamefont
  {S.}~\bibnamefont {{Zhou}}}, \bibinfo {author} {\bibfnamefont
  {S.}~\bibnamefont {{Zhou}}}, \bibinfo {author} {\bibfnamefont
  {Z.}~\bibnamefont {{Zhou}}}, \bibinfo {author} {\bibfnamefont
  {C.}~\bibnamefont {{Zhu}}}, \bibinfo {author} {\bibfnamefont
  {Q.}~\bibnamefont {{Zhu}}}, \bibinfo {author} {\bibfnamefont
  {G.}~\bibnamefont {{Zou}}}, \bibinfo {author} {\bibfnamefont
  {H.}~\bibnamefont {{Zou}}}, \bibinfo {author} {\bibfnamefont
  {Q.}~\bibnamefont {{Zhang}}}, \bibinfo {author} {\bibfnamefont {C.-Y.}\
  \bibnamefont {{Lu}}}, \bibinfo {author} {\bibfnamefont {C.-Z.}\ \bibnamefont
  {{Peng}}}, \bibinfo {author} {\bibfnamefont {X.}~\bibnamefont {{Yuan}}},
  \bibinfo {author} {\bibfnamefont {M.}~\bibnamefont {{Gong}}}, \bibinfo
  {author} {\bibfnamefont {X.}~\bibnamefont {{Zhu}}},\ and\ \bibinfo {author}
  {\bibfnamefont {J.-W.}\ \bibnamefont {{Pan}}},\ }\bibfield  {title} {\bibinfo
  {title} {{One- and two-dimensional cluster states for topological phase
  simulation and measurement-based quantum computation}},\ }\href
  {https://doi.org/10.1038/s41567-026-03179-6} {\bibfield  {journal} {\bibinfo
  {journal} {Nature Physics}\ }\textbf {\bibinfo {volume} {22}},\ \bibinfo
  {pages} {430} (\bibinfo {year} {2026})}\BibitemShut {NoStop}%
\bibitem [{\citenamefont {Pokharel}\ and\ \citenamefont
  {Lidar}(2024)}]{pokharel2024better}%
  \BibitemOpen
  \bibfield  {author} {\bibinfo {author} {\bibfnamefont {B.}~\bibnamefont
  {Pokharel}}\ and\ \bibinfo {author} {\bibfnamefont {D.~A.}\ \bibnamefont
  {Lidar}},\ }\bibfield  {title} {\bibinfo {title} {Better-than-classical
  {Grover} search via quantum error detection and suppression},\ }\href@noop {}
  {\bibfield  {journal} {\bibinfo  {journal} {npj Quantum Information}\
  }\textbf {\bibinfo {volume} {10}},\ \bibinfo {pages} {23} (\bibinfo {year}
  {2024})}\BibitemShut {NoStop}%
\bibitem [{\citenamefont {Roth}\ and\ \citenamefont
  {Peranson}(1999)}]{roth1999redesign}%
  \BibitemOpen
  \bibfield  {author} {\bibinfo {author} {\bibfnamefont {A.~E.}\ \bibnamefont
  {Roth}}\ and\ \bibinfo {author} {\bibfnamefont {E.}~\bibnamefont
  {Peranson}},\ }\bibfield  {title} {\bibinfo {title} {The redesign of the
  matching market for {American} physicians: Some engineering aspects of
  economic design},\ }\href@noop {} {\bibfield  {journal} {\bibinfo  {journal}
  {American economic review}\ }\textbf {\bibinfo {volume} {89}},\ \bibinfo
  {pages} {748} (\bibinfo {year} {1999})}\BibitemShut {NoStop}%
\bibitem [{\citenamefont {{National Resident Matching
  Program}}(2026)}]{nrmp2026report}%
  \BibitemOpen
  \bibfield  {author} {\bibinfo {author} {\bibnamefont {{National Resident
  Matching Program}}},\ }\href@noop {} {\bibinfo {title} {{NRMP} releases
  results of the 2026 main residency match for more than 38,000 future
  residents}},\ \bibinfo {howpublished}
  {\url{https://www.nrmp.org/about/news/2026/03/nrmp-releases-results-of-the-2026-main-residency-match-for-more-than-38000-future-residents/}}
  (\bibinfo {year} {2026}),\ \bibinfo {note} {published March 20,
  2026}\BibitemShut {NoStop}%
\bibitem [{SM()}]{SM}%
  \BibitemOpen
  \href@noop {} {}\bibinfo {note} {See Supplemental Material at [URL will be
  inserted by publisher] for additional theoretical and experimental details,
  which includes
  Refs.~\cite{guo2024experimental,McKay2017EfficientZGates,rol2020time}.}\BibitemShut
  {Stop}%
\bibitem [{\citenamefont {Dawson}\ and\ \citenamefont
  {Nielsen}(2006)}]{Dawson2006SolovayKitaev}%
  \BibitemOpen
  \bibfield  {author} {\bibinfo {author} {\bibfnamefont {C.~M.}\ \bibnamefont
  {Dawson}}\ and\ \bibinfo {author} {\bibfnamefont {M.~A.}\ \bibnamefont
  {Nielsen}},\ }\bibfield  {title} {\bibinfo {title} {The {Solovay-Kitaev}
  algorithm},\ }\href@noop {} {\bibfield  {journal} {\bibinfo  {journal}
  {Quantum Information and Computation}\ }\textbf {\bibinfo {volume} {6}},\
  \bibinfo {pages} {81} (\bibinfo {year} {2006})},\ \Eprint
  {https://arxiv.org/abs/quant-ph/0505030} {arXiv:quant-ph/0505030}
  \BibitemShut {NoStop}%
\bibitem [{\citenamefont {Nie}\ \emph {et~al.}(2024)\citenamefont {Nie},
  \citenamefont {Zi},\ and\ \citenamefont
  {Sun}}]{Nie2024QuantumCircuitToffoli}%
  \BibitemOpen
  \bibfield  {author} {\bibinfo {author} {\bibfnamefont {J.}~\bibnamefont
  {Nie}}, \bibinfo {author} {\bibfnamefont {W.}~\bibnamefont {Zi}},\ and\
  \bibinfo {author} {\bibfnamefont {X.}~\bibnamefont {Sun}},\ }\href
  {https://arxiv.org/abs/2402.05053} {\bibinfo {title} {Quantum circuit for
  multi-qubit {Toffoli} gate with optimal resource}} (\bibinfo {year} {2024}),\
  \Eprint {https://arxiv.org/abs/2402.05053} {arXiv:2402.05053 [quant-ph]}
  \BibitemShut {NoStop}%
\bibitem [{\citenamefont {Guo}\ \emph {et~al.}(2024)\citenamefont {Guo},
  \citenamefont {Sun}, \citenamefont {Qian}, \citenamefont {Gong},
  \citenamefont {Zhang}, \citenamefont {Chen}, \citenamefont {Ye},
  \citenamefont {Wu}, \citenamefont {Cao}, \citenamefont {Liu} \emph
  {et~al.}}]{guo2024experimental}%
  \BibitemOpen
  \bibfield  {author} {\bibinfo {author} {\bibfnamefont {S.}~\bibnamefont
  {Guo}}, \bibinfo {author} {\bibfnamefont {J.}~\bibnamefont {Sun}}, \bibinfo
  {author} {\bibfnamefont {H.}~\bibnamefont {Qian}}, \bibinfo {author}
  {\bibfnamefont {M.}~\bibnamefont {Gong}}, \bibinfo {author} {\bibfnamefont
  {Y.}~\bibnamefont {Zhang}}, \bibinfo {author} {\bibfnamefont
  {F.}~\bibnamefont {Chen}}, \bibinfo {author} {\bibfnamefont {Y.}~\bibnamefont
  {Ye}}, \bibinfo {author} {\bibfnamefont {Y.}~\bibnamefont {Wu}}, \bibinfo
  {author} {\bibfnamefont {S.}~\bibnamefont {Cao}}, \bibinfo {author}
  {\bibfnamefont {K.}~\bibnamefont {Liu}}, \emph {et~al.},\ }\bibfield  {title}
  {\bibinfo {title} {Experimental quantum computational chemistry with
  optimized unitary coupled cluster ansatz},\ }\href@noop {} {\bibfield
  {journal} {\bibinfo  {journal} {Nature Physics}\ }\textbf {\bibinfo {volume}
  {20}},\ \bibinfo {pages} {1240} (\bibinfo {year} {2024})}\BibitemShut
  {NoStop}%
\bibitem [{\citenamefont {McKay}\ \emph {et~al.}(2017)\citenamefont {McKay},
  \citenamefont {Wood}, \citenamefont {Sheldon}, \citenamefont {Chow},\ and\
  \citenamefont {Gambetta}}]{McKay2017EfficientZGates}%
  \BibitemOpen
  \bibfield  {author} {\bibinfo {author} {\bibfnamefont {D.~C.}\ \bibnamefont
  {McKay}}, \bibinfo {author} {\bibfnamefont {C.~J.}\ \bibnamefont {Wood}},
  \bibinfo {author} {\bibfnamefont {S.}~\bibnamefont {Sheldon}}, \bibinfo
  {author} {\bibfnamefont {J.~M.}\ \bibnamefont {Chow}},\ and\ \bibinfo
  {author} {\bibfnamefont {J.~M.}\ \bibnamefont {Gambetta}},\ }\bibfield
  {title} {\bibinfo {title} {Efficient {Z} gates for quantum computing},\
  }\href {https://doi.org/10.1103/PhysRevA.96.022330} {\bibfield  {journal}
  {\bibinfo  {journal} {Physical Review A}\ }\textbf {\bibinfo {volume} {96}},\
  \bibinfo {pages} {022330} (\bibinfo {year} {2017})}\BibitemShut {NoStop}%
\bibitem [{\citenamefont {Rol}\ \emph {et~al.}(2020)\citenamefont {Rol},
  \citenamefont {Ciorciaro}, \citenamefont {Malinowski}, \citenamefont
  {Tarasinski}, \citenamefont {Sagastizabal}, \citenamefont {Bultink},
  \citenamefont {Salathe}, \citenamefont {Haandb{\ae}k}, \citenamefont
  {Sedivy},\ and\ \citenamefont {DiCarlo}}]{rol2020time}%
  \BibitemOpen
  \bibfield  {author} {\bibinfo {author} {\bibfnamefont {M.~A.}\ \bibnamefont
  {Rol}}, \bibinfo {author} {\bibfnamefont {L.}~\bibnamefont {Ciorciaro}},
  \bibinfo {author} {\bibfnamefont {F.~K.}\ \bibnamefont {Malinowski}},
  \bibinfo {author} {\bibfnamefont {B.~M.}\ \bibnamefont {Tarasinski}},
  \bibinfo {author} {\bibfnamefont {R.~E.}\ \bibnamefont {Sagastizabal}},
  \bibinfo {author} {\bibfnamefont {C.~C.}\ \bibnamefont {Bultink}}, \bibinfo
  {author} {\bibfnamefont {Y.}~\bibnamefont {Salathe}}, \bibinfo {author}
  {\bibfnamefont {N.}~\bibnamefont {Haandb{\ae}k}}, \bibinfo {author}
  {\bibfnamefont {J.}~\bibnamefont {Sedivy}},\ and\ \bibinfo {author}
  {\bibfnamefont {L.}~\bibnamefont {DiCarlo}},\ }\bibfield  {title} {\bibinfo
  {title} {Time-domain characterization and correction of on-chip distortion of
  control pulses in a quantum processor},\ }\href@noop {} {\bibfield  {journal}
  {\bibinfo  {journal} {Applied Physics Letters}\ }\textbf {\bibinfo {volume}
  {116}},\ \bibinfo {pages} {054001} (\bibinfo {year} {2020})}\BibitemShut
  {NoStop}%
\end{thebibliography}
\end{document}


\maketitle
\begin{center}
\small
\textsuperscript{*}These authors contributed equally to this work.

\textsuperscript{\dag}Corresponding author:
\href{mailto:sunxiaoming@ict.ac.cn}{sunxiaoming@ict.ac.cn}

\textsuperscript{\ddag}Corresponding author:
\href{mailto:chenzha@hfnl.cn}{chenzha@hfnl.cn}
\end{center}

\section{Implementation of the DTQW Circuit}

The deterministic discrete-time quantum-walk (DTQW) search protocol follows Ref.~\cite{Peng2024DeterministicDTQW}, which gives the operator definitions, parameter schedule, and circuit construction for complete bipartite graphs. Here we record the encoding and synthesis conventions used to generate the analyzed circuits, including the balanced full-space instances used in the resource estimates and the special fixed-ratio instances used experimentally.

The walk state is written as $\ket{v,u}$, where $v$ labels the current vertex and $u$ labels the outgoing edge direction. For a complete bipartite graph, each outgoing edge from $v$ is uniquely specified by its endpoint in the opposite partition, so the edge register can be encoded by the opposite-side vertex label. For the balanced instances used in the resource-scaling analysis, $N_1=N_2=2^n$, so the total graph size is $N=N_1+N_2=2^{n+1}$.

Each walk step implements
\[
U(\alpha_j,\beta_j)=S C(\alpha_j) Q(\beta_j),
\]
with the phase parameters $\{\alpha_j,\beta_j\}$ chosen according to the deterministic schedule of Ref.~\cite{Peng2024DeterministicDTQW}.
The oracle and coin follow the definitions in the main text and Ref.~\cite{Peng2024DeterministicDTQW}. For the general full-space circuits and resource estimation, the marked vertex is chosen to be the all-one bit string in the marked partition. Therefore, $Q(\beta_j)$ is implemented directly as a parametrized multi-controlled phase on the all-one current-vertex pattern. For an arbitrary marked bit string, the same construction applies after conjugating the corresponding zero-valued control qubits by Pauli-$X$ gates.
The coin $C(\alpha_j)$ is implemented by conjugating a multi-controlled phase with Hadamard gates on the edge register, realizing the generalized Grover coin about the uniform outgoing-edge state. All multi-controlled phase rotations appearing in the oracle and coin blocks are synthesized using the construction of Corollary~2 in Ref.~\cite{Nie2024QuantumCircuitToffoli}, which gives logarithmic depth with one ancillary qubit. This ancillary qubit is omitted from Fig.~\ref{fig:dtqw-block-implementation}, which is intended to show the logical block structure rather than the gate-level decomposition of each multi-controlled phase.
The shift $S$ maps $\ket{v,u}$ to $\ket{u,v}$. In the binary encoding, the vertex register consists of a partition bit and a within-partition label, while the edge register stores the opposite-side label. Thus, if the current vertex is encoded as $(p,x)$ and the outgoing edge points to $(1-p,y)$, then the shifted state should have current vertex $(1-p,y)$ and edge direction pointing back to $(p,x)$. This is implemented by swapping the two label registers, $x\leftrightarrow y$, and flipping the partition bit, $p\mapsto 1-p$.
In Fig.~\ref{fig:dtqw-block-implementation}, $P(\theta)$ denotes the multi-controlled phase operation that multiplies the selected computational-basis state by $e^{i\theta}$.
Figure~\ref{fig:dtqw-block-implementation} summarizes the block-level circuit implementation of the oracle, coin, and shift operators used in the generated DTQW circuits.

\begin{figure}[htbp]
    \centering
    \resizebox{0.82\linewidth}{!}{\input{dtqw_block_implementation}}
    \caption{Block-level implementation of the DTQW walk-step components. (a) The oracle is a multi-controlled phase on the all-one marked current-vertex pattern. (b) The coin is implemented as a $P(-\alpha_j)$ phase about the uniform outgoing-edge state by conjugating a multi-controlled phase with Hadamards on the edge register. (c) The shift swaps the within-partition vertex label with the edge label and flips the partition bit, realizing $S\ket{v,u}=\ket{u,v}$.}
    \label{fig:dtqw-block-implementation}
\end{figure}

For the special case $n_1/N_1=1/4$, the simplification to a fixed sequence of operators follows from the parameter schedule in Ref.~\cite{Peng2024DeterministicDTQW}. Since $n_1/N_1=1/4$ gives $\omega=2\arcsin\sqrt{n_1/N_1}=\pi/3$, the generated circuit uses the fixed block sequence
\[
Q(-\pi)\,S\,Q(-\pi)\,C(\pi)\,S\,Q(-3\pi/2)\,C(\pi),
\]
after the initial Hadamard preparation. The coin $C$ and shift $S$ use the same logical constructions as above. To keep the marked fraction fixed as $N_1=2^n$ increases, we choose the marked set $M\subset V_1$ to be the vertices whose binary labels start with the prefix $11$, i.e., $\ket{11x_3\cdots x_n}$, which gives $n_1=2^{n-2}$. The oracle $Q$ therefore reduces to a doubly controlled phase conditioned on the partition bit $p=1$ and the first two vertex-label bits being $11$, equivalently the pattern $p x_1 x_2=111$.

For the reduced-subspace circuits, we use the four-dimensional invariant subspace for the case $M\subset V_1$ derived in Sec.~IV of Ref.~\cite{Peng2024DeterministicDTQW}. The basis states $\{\ket{\phi_1},\ket{\phi_2},\ket{\phi_3},\ket{\phi_4}\}$ defined in the main text are encoded into two computational qubits in that order. In this representation, $S$ and $Q(\beta)$ are implemented directly from their subspace matrices. The coin matrix $C(\alpha)$ is instead synthesized using the decomposition from Ref.~\cite{Peng2024DeterministicDTQW},
\[
C(\alpha)\sim A(\pi/2)R(\alpha)A(-\pi/2),
\]
where the equivalence is up to a global phase. We first construct circuits for the elementary two-qubit operations $A(\theta)$ and $R(\theta)$ and then compose them to implement $C(\alpha)$, with the usual right-to-left operator ordering.

\section{Resource-Scaling and Fidelity-Estimate Details}
\label{sec:resource-estimate-details}

\begin{figure}[htbp]
    \centering
    \includegraphics[width=0.70\textwidth]{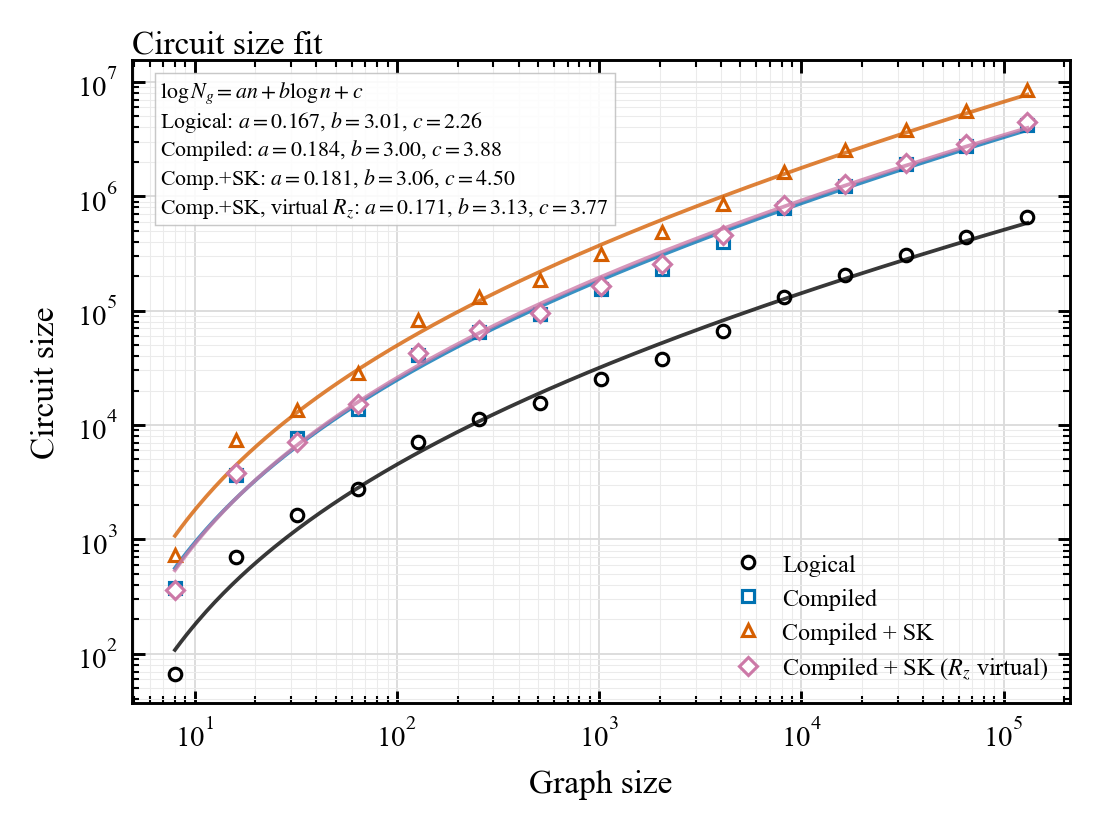}
    \caption{Circuit-size scaling and fitted gate-count model for the general full-space circuits. Markers show explicitly generated circuit sizes, and solid curves show the fitted curves with coefficients listed in the panel. The curve for circuits obtained by Solovay--Kitaev (SK) synthesis with virtual $R_z$ gates treated as frame updates is the gate-count convention used for panel (c) of Fig.~2 of the main text.}
    \label{fig:size-fit-virtual-rz}
\end{figure}

The main text defines the logical, hardware-compiled, and SK-synthesized circuit representations used in Fig.~2 of the main text. The SK-synthesized circuit is obtained by applying Solovay--Kitaev (SK) synthesis to the hardware-compiled circuit, excluding quantum error-correction overhead. Here we clarify how $R_z$ rotations are counted. On superconducting hardware, an $R_z$ rotation can be implemented virtually by updating the phase reference of subsequent microwave pulses, without applying an additional physical pulse~\cite{McKay2017EfficientZGates}. Panels (a) and (b) of Fig.~2 of the main text retain these virtual $R_z$ operations because they report the depth and size of each generated circuit representation. Panel (c), by contrast, estimates accumulated error from physically executed gates, so virtual $R_z$ rotations are excluded from its effective SK-synthesized gate count. Figure~\ref{fig:size-fit-virtual-rz} shows both counting conventions explicitly. After excluding virtual $R_z$ rotations, the SK-synthesized size curve retains the same smooth scaling trend as the other representations, supporting the use of the same empirical fit for the fidelity estimate.

For the extrapolated curves, we fit the logical gate count and the SK-synthesized gate count after excluding virtual $R_z$ rotations to
\[
\log N_g = a n + b\log n + c,
\]
where $N_g$ is the gate count for the circuit representation considered, $N=2^{n+1}$, and $\log$ denotes the natural logarithm. The fitted coefficients are $(a,b,c)=(0.166736,3.011205,2.262372)$ for the logical circuit and $(a,b,c)=(0.170708,3.132927,3.774250)$ for the SK-synthesized circuit after excluding virtual $R_z$ rotations. These fits are also displayed in Fig.~\ref{fig:size-fit-virtual-rz}. Under the product model used in the main text, the fitted total circuit fidelity is
\[
F_{\mathrm{tot}}(n)=f_g^{\widehat{N}_g(n)},
\]
where $\widehat{N}_g(n)$ denotes the fitted gate count. Solving for the required gate infidelity at a specified value of $F_{\mathrm{tot}}$ gives
\[
1-f_g = 1-\exp\!\left(\frac{\log F_{\mathrm{tot}}}{\widehat{N}_g(n)}\right).
\]
The curve in panel (c) of Fig.~2 of the main text uses $F_{\mathrm{tot}}=0.5$.

\begin{figure}[htbp]
    \centering
    \includegraphics[width=0.70\textwidth]{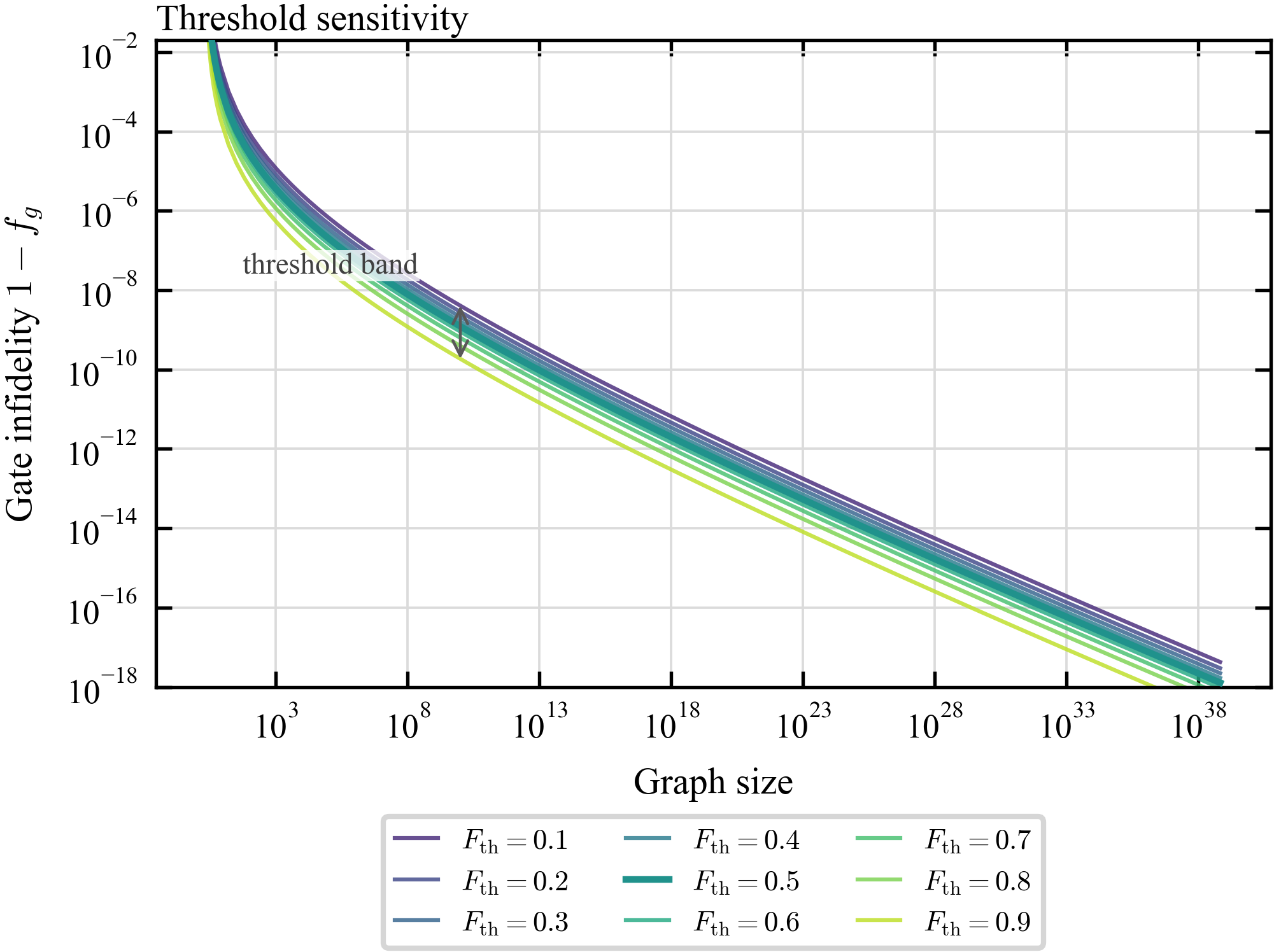}
    \caption{Sensitivity of the gate-infidelity requirement for circuits obtained by Solovay--Kitaev (SK) synthesis to the selected total circuit fidelity. Curves use the same fitted gate-count model as panel (c) of Fig.~2 of the main text, with $F_{\mathrm{tot}}$ varied from $0.1$ to $0.9$. This choice produces a constant vertical rescaling of the required infidelity and does not change the graph-size scaling.}
    \label{fig:threshold-sensitivity}
\end{figure}

The choice $F_{\mathrm{tot}}=0.5$ is used as an order-unity circuit-fidelity criterion rather than a sharp algorithmic threshold. In the small-infidelity regime,
$1-f_g\simeq -\log(F_{\mathrm{tot}})/\widehat{N}_g(n)$, so changing the selected value of $F_{\mathrm{tot}}$ within the same order mainly rescales the vertical position of the curve while leaving its graph-size dependence fixed. Figure~\ref{fig:threshold-sensitivity} shows this explicitly for $F_{\mathrm{tot}}=0.1$--$0.9$.

For the current-hardware estimate, the SK-synthesized circuit is also counted after separating nonvirtual one-qubit gates and CZ gates. Their fitted counts, $\widehat{N}_{1q}(n)$ and $\widehat{N}_{\mathrm{CZ}}(n)$, are combined with the median calibrated \textit{Zuchongzhi 3.1} fidelities through the same product model,
\[
F_{\mathrm{HW}}(n)
= f_{1q}^{\widehat{N}_{1q}(n)}
  f_{\mathrm{CZ}}^{\widehat{N}_{\mathrm{CZ}}(n)}.
\]
The vertical marker in panel (c) of Fig.~2 of the main text is the graph size at which this fitted $F_{\mathrm{HW}}(n)$ crosses the selected value $F_{\mathrm{tot}}=0.5$. To express the heterogeneous one- and two-qubit errors on the same per-gate scale as the preceding model, we define an effective fidelity $f_{g,\mathrm{eff}}(n)$ by
\[
F_{\mathrm{HW}}(n)
=f_{g,\mathrm{eff}}(n)^{\widehat{N}_{1q}(n)+\widehat{N}_{\mathrm{CZ}}(n)}.
\]
Taking the logarithm of this definition and substituting the product expression for $F_{\mathrm{HW}}(n)$ gives a gate-count-weighted geometric mean of $f_{1q}$ and $f_{\mathrm{CZ}}$. Therefore, the gray hardware curve in the same panel reports the effective infidelity
\[
1-f_{g,\mathrm{eff}}(n)=1-\exp\!\left[
\frac{\widehat{N}_{1q}(n)\log f_{1q}
      +\widehat{N}_{\mathrm{CZ}}(n)\log f_{\mathrm{CZ}}}
     {\widehat{N}_{1q}(n)+\widehat{N}_{\mathrm{CZ}}(n)}
\right].
\]

\section{Quantum Processor and Calibration}

\begin{figure}[htbp]
    \centering
    \includegraphics[width=0.6\textwidth]{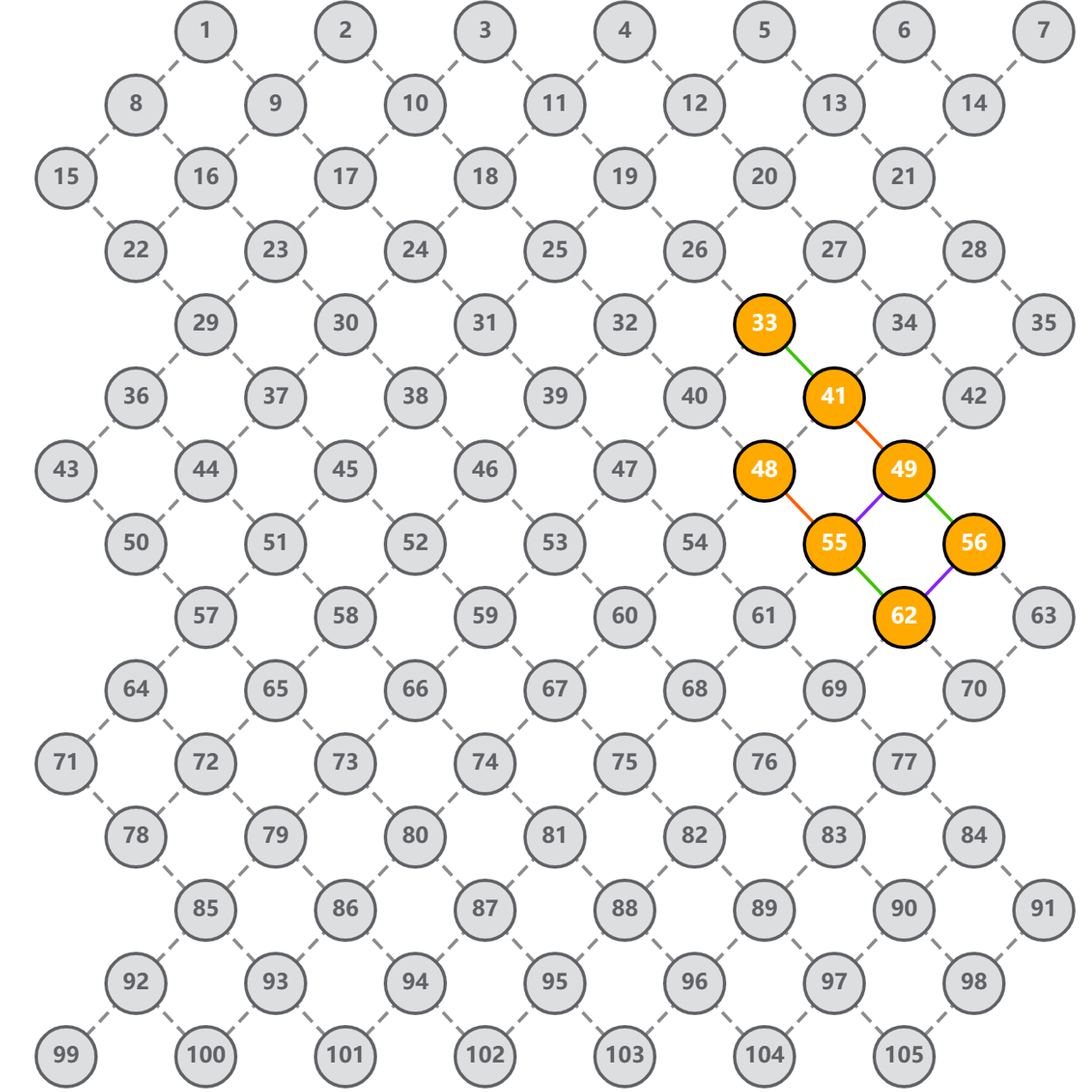}
    \caption{The 7-qubit topology utilized in this work, selected from the 105 qubits of the \textit{Zuchongzhi 3.1} processor.}
    \label{fig:topo}
\end{figure}

The processor used in this work is the \textit{Zuchongzhi 3.1} superconducting quantum processor \cite{jiang2025generation95qubitgenuineentanglement}, which comprises 105 transmon qubits and 182 tunable couplers. The full topology of the processor is illustrated in Fig.~\ref{fig:topo}. In our quantum-walk search experiment, we utilized a subset of up to 7 qubits and 7 couplers.

We optimized the required qubits and couplers, including single-qubit gate fidelity, CZ gate fidelity, readout fidelity, and the isolation of unused qubits by turning off their respective couplings. Both single-qubit and two-qubit gates have a duration of 44~ns, while the readout time is 1300~ns. Detailed system parameters are summarized in Table~\ref{tab:qubit_params}.

According to the requirements of the experimental circuits, we grouped the CZ gates that need to be executed in parallel into specific patterns. We individually calibrated and optimized the CZ gate fidelity within each pattern. Notably, due to the structure of the experimental quantum circuits, a specific CZ gate may appear in multiple patterns. The CZ gate fidelities reported in Table~\ref{tab:qubit_params} represent the minimum values observed across all patterns in which the gate participates.

\begin{table}[htbp]
\centering
\caption{Detailed parameters for the 7 qubits and their corresponding CZ gates.}
\label{tab:qubit_params}
\resizebox{\columnwidth}{!}{
\begin{tabular}{@{}lccccccc@{}}
\toprule
Qubit & $q_{33}$ & $q_{41}$ & $q_{48}$ & $q_{49}$ & $q_{55}$ & $q_{56}$ & $q_{62}$ \\ \midrule
$f_{01}$ (GHz) & 5.5140 & 5.3660 & 5.4010 & 5.4480 & 5.4160 & 5.3550 & 5.4832 \\
$\eta$ (MHz) & -211.1 & -213.8 & -216.2 & -214.6 & -213.1 & -217.2 & -210.6 \\
$\pi/2$ fidelity (XEB) & 99.930\% & 99.926\% & 99.931\% & 99.903\% & 99.931\% & 99.929\% & 99.909\% \\ 
Readout fid. & 99.64\% & 99.53\% & 99.64\% & 99.53\% & 99.57\% & 99.27\% & 99.63\% \\
$T_1$ ($\mu$s) & 38.0 & 43.2 & 63.2 & 57.3 & 48.3 & 44.5 & 42.4\\
$T_2^*$ ($\mu$s) & 16.8 & 7.2 & 5.1 & 12.0 & 13.0 & 8.5 & 4.4 \\ \midrule
Connected to & $q_{41}$ & $q_{49}$ & $q_{55}$ & $q_{55}$ & $q_{62}$ & $q_{49}$ & $q_{56}$ \\ 
CZ fidelity (XEB) & 99.60\% & 99.70\% & 99.29\% & 99.82\% & 99.46\% & 99.46\% & 99.70\% \\ \midrule
\bottomrule
\end{tabular}
}
\end{table}

\section{Correction to the Dynamical-Phase Correction Method}

The implementation of CZ gates requires fast frequency tuning of both qubits and couplers via Z-pulses. These pulses are often distorted and typically suffer from significant transients (tails), which can negatively impact subsequent gate operations. While some approaches utilize frequency-control pulse distortion corrections \cite{rol2020time}, accurately measuring the exact shape of the tail in an experimental setting is challenging due to the complex influences of cable configurations and chip design. 

In our experiment, the impact of these tails is particularly pronounced due to the depth of the circuits and the high number of CZ gates. To mitigate this, we developed a simple and effective transient correction method.

\begin{figure}[htbp]
    \centering
    \includegraphics[width=0.85\textwidth]{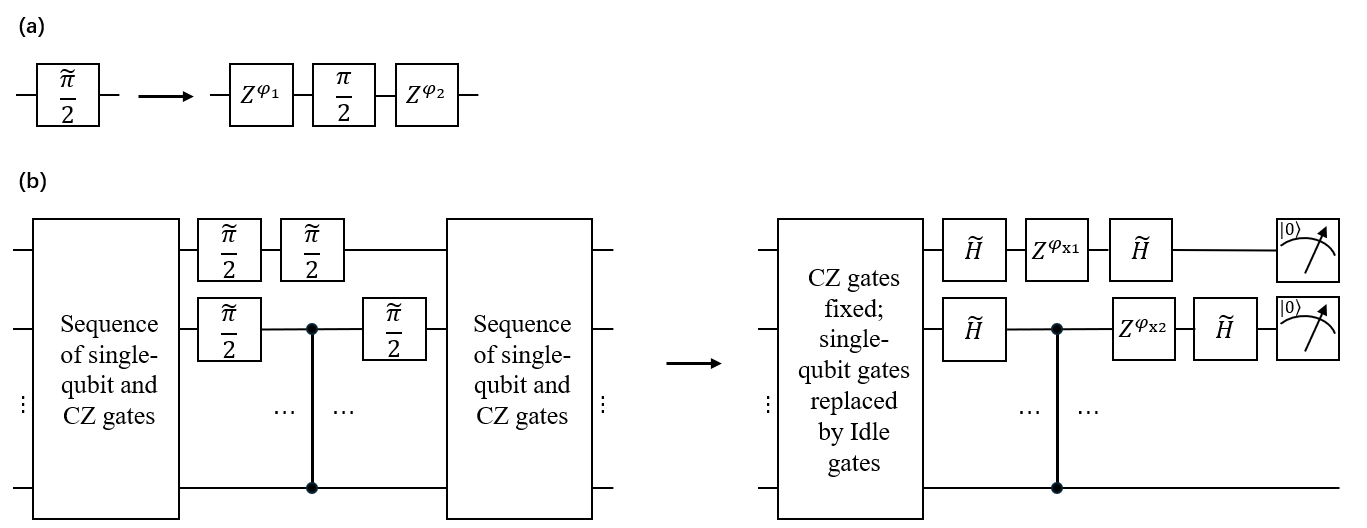}
    \caption{Measurement of dynamical phases induced by CZ tails. (a) A $\pi/2$ pulse with a small transient is equivalent to a perfect pulse with $Z$-rotations. (b) To measure the phase, preceding gates are replaced by Idles while CZ gates remain. We scan the $Z$ angle between two $H$ gates; if the angle that yields the maximum $\ket{0}$ probability is $\varphi$, the dynamical phase is $-\varphi$.}
    \label{fig:dyn_phase}
\end{figure}

As shown in Fig.~\ref{fig:dyn_phase}(a), a $\pi/2$ pulse affected by a Z-tail can be modeled as an ideal $\pi/2$ pulse flanked by small Z-rotations (provided the tail is not excessively large). We refer to these rotation angles as the dynamical phase. In a complete circuit, the Z-rotation before the first $\pi/2$ gate and after the final $\pi/2$ gate only contribute a global phase. Therefore, our objective is to determine the Z-rotation angles between all subsequent pairs of $\pi/2$ gates.

Taking two affected $\pi/2$ gates as an example (Fig.~\ref{fig:dyn_phase}(b)):
\begin{enumerate}
    \item We replace all single-qubit gates preceding the two $\pi/2$ gates with Idle gates of equivalent duration.
    \item The CZ gates are kept in their original positions to maintain the same transient environment as the actual experiment.
    \item The two $\pi/2$ gates under test are replaced by Hadamard (H) gates.
    \item We scan the Z-rotation angle between these two gates to find the angle that maximizes the probability of the final output state being $\ket{0}$.
\end{enumerate}

All Z-rotations in our experiment are implemented as virtual Z gates. By performing this scan between all adjacent single-qubit ($\pi/2$) gates for all qubits, we obtain the required Z-rotation angles. In the final experimental circuits, we insert a Z-rotation before each $\pi/2$ gate to cancel out the dynamical phase induced by the tails.

\section{Readout Error Mitigation}

To mitigate readout errors, we characterize the single-qubit readout response of each qubit and construct a calibration matrix as the tensor product of the individual response matrices. To assess the validity of this approximation, we performed random-state measurement experiments~\cite{guo2024experimental} to characterize readout crosstalk. The results indicate that correlated readout errors are sufficiently small, justifying the use of the tensor-product approximation for our experiments.

For a single qubit $i$, the readout error can be defined by a $2 \times 2$ matrix:
\begin{equation}
\Lambda_i = \begin{bmatrix}
p(0|0) & p(0|1) \\
p(1|0) & p(1|1)
\end{bmatrix} = \begin{bmatrix}
1-\epsilon_i & \eta_i \\
\epsilon_i & 1-\eta_i
\end{bmatrix}
\end{equation}
where $p(x|y)$ represents the probability of measuring state $\ket{x}$ given that the state was prepared as $\ket{y}$. Here, $\epsilon_i$ and $\eta_i$ denote the bit-flip error rates for $\ket{0} \to \ket{1}$ and $\ket{1} \to \ket{0}$, respectively.

For an $n$-qubit system, the total noise matrix $\Lambda$ is the tensor product of the individual qubit matrices:
\begin{equation}
\Lambda = \Lambda_1 \otimes \Lambda_2 \otimes \dots \otimes \Lambda_n
\end{equation}
Given a noisy measured probability distribution $\vec{p}$, the corrected distribution $\vec{p'}$ can be calculated using the inverse of the noise matrix:
\begin{equation}
\vec{p'} = \Lambda^{-1} \vec{p}
\end{equation}

\section{Quantum State Tomography}
\begin{figure}[htbp]
    \centering
    \includegraphics[width=0.85\textwidth]{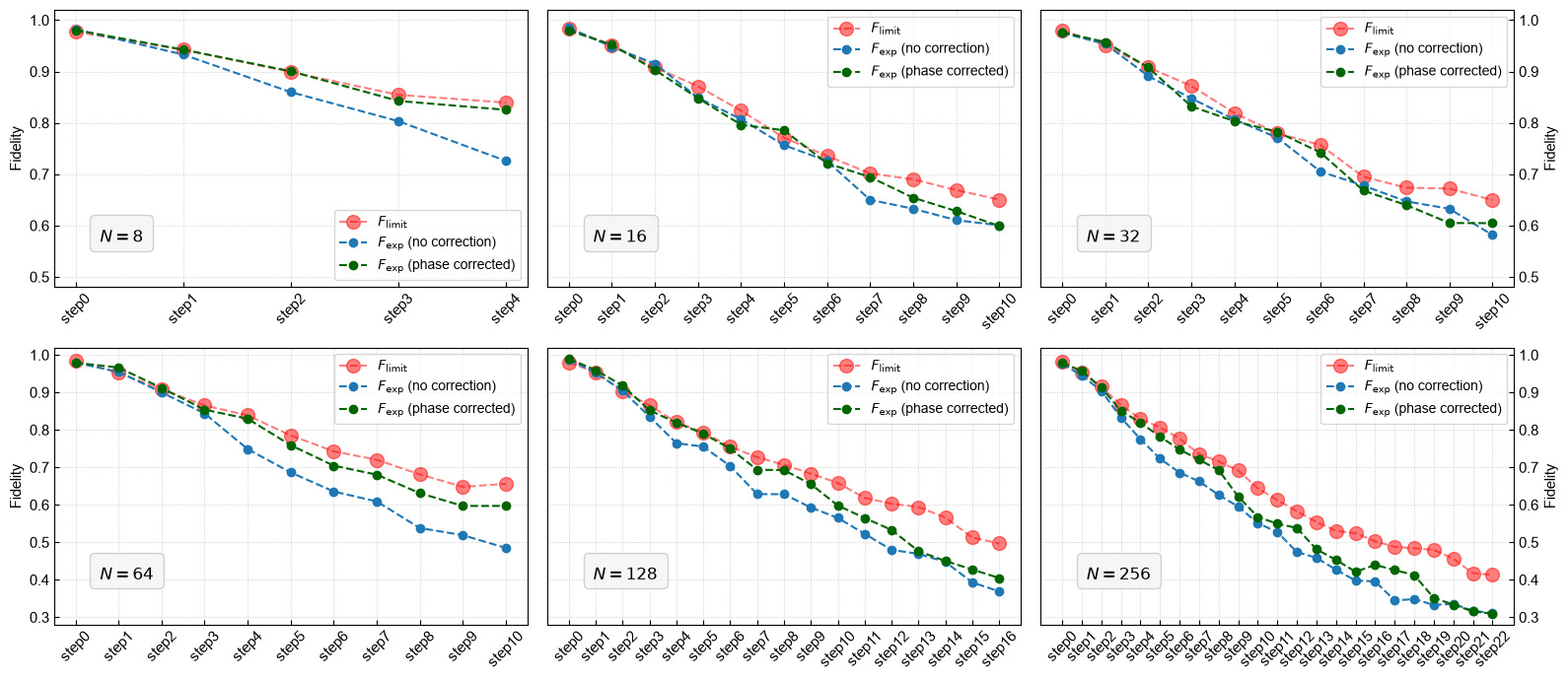}
    \caption{Step-by-step quantum state tomography results for all six subspace circuits. The purity-limited fidelity upper bound $F_\text{limit}$ (red) is compared with the experimentally reconstructed state fidelity $F_\text{exp}$ obtained without (blue) and with (green) dynamical-phase correction. Each point corresponds to the output state after one walk step.}
    \label{fig:tomo_fid}
\end{figure}

We further employed quantum state tomography (QST) to determine whether experimental errors originate from coherent or incoherent sources. We performed QST on the output of each step for all subspace circuits, using purity $P= \text{tr}(\rho^2)$ and fidelity $F(\rho, \sigma) = \left( \text{tr} \sqrt{\sqrt{\rho} \sigma \sqrt{\rho}} \right)^2$ to evaluate performance. Here, $\rho$ is the reconstructed experimental density matrix and $\sigma$ is the corresponding ideal output state. We define $F_{\text{limit}}$ as the purity-limited fidelity upper bound:
\begin{equation}
    F_{\text{limit}} = \frac{1 + \sqrt{(2^n - 1)(P \cdot 2^n - 1)}}{2^n}
    \label{eq:f_limit}
\end{equation}
where $n$ denotes the number of qubits employed in the circuit, such that $d = 2^n$ represents the dimension of the Hilbert space. This upper bound $F_{\text{limit}}$, derived from the state purity $P$, characterizes the maximum attainable fidelity in the presence of purely incoherent noise.
We found that the experimental state fidelity closely aligns with the theoretical bound determined by purity, as shown in Fig.~\ref{fig:tomo_fid}. Additionally, we analyzed the purity ($P$), fidelity ($F$), and the purity-limited fidelity upper bound ($F_{\text{limit}}$) for the $N = 8$ general case and the $N =8 $ special case (5 qubits). For the general case, the measured values are $P = 31.2\%$, $F = 41.6\%$, and $F_{\text{limit}} = 55.2\%$, respectively. For the special case, these values are $P = 27.4\%$, $F = 45.7\%$, and $F_{\text{limit}} = 51.6\%$. We observed that the experimental fidelity is slightly lower than the corresponding $F_{\text{limit}}$, which we attribute to coherent control errors introduced by the increased qubit count, such as control pulse crosstalk. Our analysis suggests that incoherent noise remains the dominant limiting factor for the observed success probabilities. While coherent crosstalk is present, the purity-limited fidelity bound $F_{\text{limit}}$ sets the primary ceiling for our current hardware performance. Future improvements in the $T_2$ of quantum processors will significantly benefit our algorithm, enabling higher search success probabilities.

\bibliographystyle{plain}